\documentclass[aps,pre,amsmath,amsfonts,floatfix,superscriptaddress,nofootinbib]{revtex4}

\usepackage{t1enc,latexsym,amsfonts, amssymb, amsmath,empheq,amsthm,bm}
\usepackage{graphicx}
\usepackage{stmaryrd}

\usepackage{mathrsfs}
\usepackage[usenames,dvipsnames]{xcolor}
\usepackage{color}
\usepackage{multirow}
\usepackage{bbm}
\usepackage{commath}

\usepackage{tikz}
\usetikzlibrary{shapes,arrows,positioning,calc}
\usepackage{hyperref}

\usepackage[french,english]{babel}
\usepackage[utf8x]{inputenc}
\usepackage[T1]{fontenc}
\usepackage{times}
\usepackage{url}

\definecolor{darkgreen}{rgb}{0,0.6,0}
\definecolor{darkblue}{rgb}{0,0,0.6}
\definecolor{darkred}{rgb}{0.6,0,0}
\definecolor{darkpurple}{rgb}{0.5,0,0.5}

\hypersetup{
bookmarksopen=true,
pdftitle="Out-of-equilibrium dynamical equations of infinite-dimensional particle systems -- II. The anisotropic case under shear strain",
pdfauthor="Agoritsas-Maimbourg-Zamponi", 
pdftoolbar=false, % toolbar hidden
pdfstartview={FitH},		% fits the width of the page to the window
pdfmenubar=true,			%menubar shown
pdfhighlight=/O,			%effect of clicking on a link
colorlinks=true,			%couleurs sur les liens hypertextes
urlcolor=darkblue,
citecolor=darkblue,		%couleur des liens pour les citations
linkcolor=darkpurple,	%darkred %couleur des liens hypertextes internes
}

\let\a=\alpha \let\b=\beta \let\g=\gamma \let\d=\delta
 \let\z=\zeta  \let\k=\kappa
\let\l=\lambda \let\m=\mu \let\n=\nu \let\x=\xi 
\let\s=\sigma  \let\f=\varphi 
   \let\G=\Gamma
\let\D=\Delta  \let\X=\Xi 
   
 \let\r=\rho \let\th=\theta \let\io=\infty

\def\ie{{\textit{i.e.} }}\def\eg{{\textit{e.g.} }}

\def\PP{{\cal P}}\def\MM{{\cal M}} 
\def\CC{{\cal C}} 
 
\def\cR{{\cal R}}  \def\OO{{\cal O}}
\def\DD{{\cal D}}\def\GG{{\cal G}}

\def\xx{{\bf x}}  

\def\uu{{\bf u}}
\def\vv{{\bf w}}

\def\ul{\underline}

\def\redv{\bar v}

\def\rr{\mathbf{r}}
\def\RR{\mathbf{R}}
\def\xx{\mathbf{x}}
\def\de{\mathrm d}
\def\Lap{\nabla^2}

\def\Bi{\b_0}

\def\to{\rightarrow} \def\la{\left\langle} \def\ra{\right\rangle}

\newcommand{\beq}{\begin{equation}} \newcommand{\eeq}{\end{equation}}
\newcommand{\wh}{\widehat} 
\newcommand{\Tr}{\text{Tr}}

\newcommand{\afunc}[1]{\operatorname{\mathsf{#1}}}
\def\DE{\afunc{D}}
\def\AE{\afunc{A}}

\newcommand{\argc}[1]{\left[#1\right]}
\newcommand{\arga}[1]{\left\lbrace #1\right\rbrace }
\newcommand{\argp}[1]{\left(#1\right)}
\newcommand{\valabs}[1]{\vert #1\vert}
\newcommand{\moy}[1]{\left\langle  #1 \right\rangle }

\begin{document}

\title{
Out-of-equilibrium dynamical equations of infinite-dimensional particle systems
\\
II.~The anisotropic case under shear strain
} 

%_____________________________________________________________

\author{Elisabeth Agoritsas}
\affiliation{Institute of Physics, EPFL, CH-1015 Lausanne, Switzerland}

\author{Thibaud Maimbourg}
\affiliation{LPTMS, CNRS, Université Paris-Sud, Université Paris-Saclay, 91405 Orsay, France}
\affiliation{The Abdus Salam International Centre for Theoretical Physics, Strada Costiera 11, 34151 Trieste, Italy}

\author{Francesco Zamponi}
\affiliation{Laboratoire de Physique de l’Ecole Normale Sup\'erieure, ENS, Universit\'e PSL, CNRS, Sorbonne Universit\'e, Universit\'e Paris-Diderot, Sorbonne Paris Cit\'e, Paris, France}

%_____________________________________________________________

\begin{abstract}

As an extension of the isotropic setting presented in the companion paper~\cite{AMZ19},
we consider the Langevin dynamics of a many-body system of pairwise interacting particles
in $d$ dimensions, submitted to an external shear strain.
We show that the anisotropy introduced by the shear strain can be simply addressed by moving into the co-shearing frame,
leading to simple dynamical mean field equations in the limit ${d\to\io}$.
The dynamics is then controlled by a single one-dimensional effective stochastic process which
depends on three distinct strain-dependent kernels
--~self-consistently determined by the process itself~--
encoding the effective restoring force, friction and noise terms due to the particle interactions.
From there one can compute dynamical observables such as particle mean-square displacements and shear stress fluctuations, and eventually aim at providing an exact ${d \to \infty}$ benchmark for liquid and glass rheology.
As an application of our results, we derive dynamically the `state-following' equations that describe the static response of a glass to a finite shear strain until it yields.

\end{abstract}

%_____________________________________________________________

\maketitle

%% Date
\begin{center}
\today
\end{center}

%_____________________________________________________________
%% Table of contents

\begin{center}
\rule{200pt}{0.5pt}
\end{center}

\tableofcontents

\begin{center}
\rule{200pt}{0.5pt}
\end{center}

%_____________________________________________________________
\section{Introduction}
\label{sec-introduction}

The response of a structurally disordered system to an external deformation is a central issue of both glass mechanics and rheology.
Nevertheless, despite decades of theoretical and numerical studies~\cite{RTV11,BDBDM17}, it still lacks an exact microscopic theory characterising at the same time solid glasses and flowing liquids.
The core difficulty lies in the out-of-equilibrium nature of the phenomena of interest, which challenges standard analytical approaches initially built for equilibrium-like situations.
On the one hand, sheared solid glasses can be studied theoretically for quasistatic deformations, 
with several recent works based on thermodynamic replica theory~\cite{YM10,Yo12} 
in the infinite-dimensional limit~\cite{YZ14,RUYZ15,RU16,UZ17}, whose predictions are remarkably relevant for finite dimensions as well~\cite{NYZ16,JY17,JUZY18}.
These descriptions however break down at a critical strain, beyond which a solid phase cannot be sustained any more: this is the so-called `yielding transition', 
whose microscopic nature is currently highly debated~\cite{JPRS16,PRS17,PPRS17,BDBDM17,OBBRT18,JY17,JUZY18}.
On the other hand, flowing dense liquids are classically modelled using for instance Mode-Coupling Theory (MCT)~\cite{BBK00,FC02,Go09,Br09,BCF12} or phenomenological elasto-plastic models~\cite{NFMB19}.
MCT is essentially the only available microscopic theory which can successfully describe a wide range of strongly-interacting particle systems.
However, it relies on uncontrolled approximations and it ceases to be valid in some regimes, \textit{e.g.}~in the vicinity of the jamming transition~\cite{IB13}.
It is thus desirable to have a fully microscopic dynamical microscopic theory that becomes exact in a well-defined limit, in order to bridge the different descriptions, between quasistatic and finite-shear-rate protocols, and between the solid and flowing phases at the yielding transition.

In the limit of infinite dimension theories often become much simpler, possibly providing exact analytical benchmarks.
Physically, increasing dimensionality progressively suppresses spatial fluctuations with respect to other features, rendering mean-field descriptions exact in the infinite-dimensional limit.
For glasses this is indeed the case, and the corresponding mean-field description at equilibrium and for quasistatics has been systematically elucidated over the last years \cite{CKPUZ17}.
A more complete discussion of the motivations for investigating liquid dynamics in infinite dimension can be found in the companion paper~\cite{AMZ19}, and references therein.
In that first paper, following previous work for equilibrium dynamics~\cite{MKZ16,Sz17}, we have derived  the exact mean-field dynamics of pairwise-interacting particle systems, in the limit of infinite dimension, in a very general setting which includes but is not restricted to equilibrium.
More specifically, we have shown how the Langevin dynamics of this many-body system can be exactly reduced in this limit to a single one-dimensional effective stochastic process described by three distinct kernels, self-consistently defined by the process itself.
This effective dynamics can then be studied both analytically and numerically.
However, the derivations presented in Ref.~\cite{AMZ19} strongly rely on the statistical isotropy of the model
--~assuming generically two isotropic friction and noise kernels, a radial pair interaction potential and a statistically isotropic initial condition~--
which is broken under an external shear strain.
This anistropy under shear thus needs to be considered carefully, and this is the goal of the present paper.

The structure of the paper is as follows.
In Sec.~\ref{sec:settings}, we start by defining the Langevin dynamics under a global shear strain, along with its corresponding change of coordinates into the co-shearing frame.
The latter is key to generalise the derivations presented in Ref.~\cite{AMZ19}: although the shear strain renders the system anisotropic, in the infinite-dimensional limit the ${(d-2)}$ directions transverse to the shear plane become effectively isotropic.
This can be invoked to radically simplify the dynamics.
We thus present in Sec.~\ref{sec:vectorial-formulation} the \emph{vectorial} formulation of the high-dimensional mean-field dynamics,
discussing first a cavity-like perturbative expansion in high dimension following the derivation presented in Refs.~\cite{Sz17,AMZ19} (Sec.~\ref{sec:vectorial-formulation-cavity}).
This allows us to write down two effective stochastic processes \emph{within the co-shearing frame}, respectively for the individual displacements of particles and for the inter-particle distances, involving three distinct time-dependent kernels self-consistently defined by these processes themselves (Sec.~\ref{sec:vectorial-formulation-effective-stoch-processes}).
Under an applied shear, these three kernels are not strictly diagonal, contrarily to the isotropic case~\cite{AMZ19}; consequently we need to examine explicitly how the breaking of the statistical isotropy is encoded in these kernels (Sec.~\ref{sec:vectorial-formulation-breaking-stat-isotropy}).
A related implication is that the correlation and response functions are also non-diagonal matrices, hence we have to distinguish their matricial dynamical equations from the evolution of their isotropic (diagonal) parts, and in particular of the mean-square-displacement (MSD) functions (Sec.~\ref{sec-dynamical-equations-vectorial}).
Exploiting further the high-dimensional limit, we show in Sec.~\ref{sec-scalar-formulation} how its mean-field description is ultimately controlled by a \emph{scalar} stochastic process involving only the isotropic parts of the kernels.
This will actually be the main result of this paper: once we go into the co-shearing frame, we recover an effective dynamics very similar to the isotropic case in Ref.~\cite{AMZ19}, 
the novelty being that the fluctuating interparticle gap remembers the initial condition within the shear plane.
As a first application of our results, we re-examine in Sec.~\ref{sec-state-following} the state-following protocol under a finite shear strain,
showing how we can recover dynamically the previous static derivations~\cite{RUYZ15}.
We finally conclude in Sec.~\ref{sec-conclusions}.

%_____________________________________________________________
\section{Setting of the problem}
\label{sec:settings}

%-----------------
\subsection{Langevin dynamics with a global shear strain}
\label{sec:settings-langevin-dyn-with-shear}

We consider a system of $N$ interacting particles in $d$ spatial dimensions,
labelled\footnote{Throughout the article we will denote:
\textit{(i)}~$\bm a$ for a vector with components $a_\m$,
\textit{(ii)}~$\hat{\bm a}=\bm a/|\bm a|$ for a unit vector with components $\hat a_\m$,
and~\textit{(iii)}~$\hat a$ for a matrix with components~$a_{\m\n}$. $\hat{1}$ is the identity matrix.} respectively by ${i=1,\dots,N}$ and ${\mu=1,\dots,d}$.
Their positions ${\ul{X}(t) = \arga{\xx_i(t) \in \Omega \subset \mathbb{R}^d}_{i=1,\dots,N}}$ at time $t$ 
belong to a region $\Omega$ of volume $|\Omega|$
--~hence at number density ${\r= N/|\Omega|}$~--
and for simplicity we assume $\Omega$ to be a cubic region with periodic boundary conditions.
Throughout our derivations, we will be interested in taking first the thermodynamic limit (${N \to \infty}$ and $|\Omega |\to\io$ at fixed $\r$) and secondly the infinite-dimensional limit (${d \to \infty}$).

In our most general setting, the interacting particles are supposed to be embedded in a fluid, and
to evolve under the following (generalised) Langevin dynamics~\cite{AMZ19}:
\begin{eqnarray}
\label{eqC3:GENLang}
 && m \ddot \xx_i(t) +  \z [\dot \xx_i(t)- \bm v_f(\xx_i(t),t) ]
 	+ \int_0^t \!\! \de s \, \G_R(t,s) \, [\dot\xx_i(s) - \bm v_f(\xx_i(s),s)] 
 	=  \bm F_i(t)  + \bm{\x}_i(t) +\bm{\l}_i(t)  \ ,
 \\
\label{eqC3:GENLang-bis}
 &&	%\quad\quad
 	\text{Gaussian noise } \arga{\bm{\xi}_i(t)}_{i=1,\dots,N}: \, \quad\quad
 	\moy{\x_{i,\m}(t)}_{\bm\xi} =0 , \qquad
 	\moy{\x_{i,\m}(t) \x_{j,\n}(s)}_{\bm\xi}
		= \delta_{ij}  \delta_{\m\n} [2 T\z  \delta(t-s) + \G_C(t,s)] \ ,
 \\
\label{eqC3:GENLang-ter}
 &&
 	\text{Interaction force and potential}: \, \quad\quad \bm F_i(t)  = -\frac{\partial V(\ul X(t))}{\partial \xx_i(t)} \ , \qquad
 	V(\ul X) = \sum_{i < j} v(\xx_i - \xx_j)
 \ .
\end{eqnarray}
The first term accounts for inertia, with $m$ the individual mass of particles.
${\bm v_f(\xx,t)}$ is the velocity of the fluid in which the particles are embedded, so that the frictional forces are proportional to the relative velocity ${\dot{\bm{x}}_i(t) - \bm v_f(\xx_i,t)}$ of the particle $i$ with respect to the fluid;
hence $\zeta$ is the (local) friction coefficient and ${\Gamma_R(t,s)}$ the (retarded) friction kernel.
The conservative force ${\bm F_i(t)}$ corresponds to the sum of the pairwise interactions with particle~$i$, encoded through a radial pair potential ${v(\xx)=v(\vert \xx \vert)}$ such as the Lennard-Jones or hard sphere potentials~\cite{hansen}.
The pair potential must be thermodynamically stable in high dimension and have a well-defined limit upon the rescaling: 
\beq
 \label{eq:potscal}
 \lim_{d \to \infty} v \argp{\ell (1+ h/d)} = \bar{v}(h)
 \ ,
\eeq
where $\ell$ is a typical interaction length, as discussed \eg~in Ref.~\cite{KMZ16}.
The Gaussian noise ${\bm\x_i(t)}$ includes two additive independent contributions to account for the noisy force due to interactions with the embedding fluid:
a white noise with coefficient ${T=\beta^{-1}}$ which is the temperature of the fluid;
a colored noise whose generic noise kernel ${\Gamma_C(t,s)}$ can describe a variety of physical situations such as active matter, random forces, etc.~\cite{AMZ19}.
Note that the brackets ${\moy{\cdots}_{\bm\xi}}$ denote the statistical average over the microscopic noise~${\bm\xi}$, that hydrodynamic interactions between particles are not considered, and that the Boltzmann constant is set to ${k_B=1}$, thus fixing the units of temperature and entropy.
At last ${\bm{\lambda}_i(t)}$ is an external field set to $0$ except when we need to define the response function.

We provide an extended discussion in the companion paper~\cite{AMZ19} about the different physical situations encompassed by this general setting, depending on the specific choices of the friction and noise kernels.
Here we want to focus on the role of the fluid velocity ${\bm v_f(\xx,t)}$, set to zero in Ref.~\cite{AMZ19}, which explicitly breaks the rotational invariance of the dynamics.
More specifically, we are interested in the role of a \emph{shear strain}\footnote{The 
generic linear case would be ${\bm v_f(\xx,t) = \hat {\dot{\g}}(t) \xx}$ with $\Tr\hat {\dot{\g}}(t)=0$ which guarantees incompressibility~\cite{BCF12}.
Although we could in principle treat this generic case, we restrict ourselves to the particular case of a shear strain where ${\dot{\gamma}_{12}(t)=\dot{\gamma}(t)}$ and the other components being zero.
As long as the number of non-zero components is finite, the infinite-dimensional analysis will be similar to the one presented here.}.
It can be simply implemented by applying a shear strain to the fluid in which the particles are immersed;
it is then assumed to result in a laminar flow of the form ${\bm v_f(\xx,t) =\dot\g(t) x_2 \hat \xx_1}$, where the flow is along direction ${\m=1}$ (${\hat \xx_1}$ is the unit vector along this direction) and its gradient is constant and directed along direction ${\mu=2}$.
Physically, ${\gamma(t)=\int_0^t \de s \dot\g(s)}$ is the accumulated shear strain along ${\hat\xx_1}$ and ${\dot\gamma(t) = \de\g/\de t}$ the corresponding shear rate.
We emphasise that this specific choice of the fluid velocity defines two special orthogonal directions (${\lbrace \hat{\xx}_1,\hat{\xx}_2 \rbrace}$ without any loss of generality), and corresponds to the minimal possible linear coupling between them.
The resulting dynamics is anisotropic in the shear plane and, as we shall see, essentially isotropic in the ${(d-2)}$ other directions.
For simplicity, we will mostly consider in what follows the case with ${m=0}$ and ${\Gamma_R \equiv 0}$, in which Eq.~\eqref{eqC3:GENLang} corresponds to the more standard dynamics~\cite{FC02,Br09,BCF12,IBS13,KCIB15}:
\beq
\label{eqC3:GENLang-shear}
\begin{split}
 & \z [\dot \xx_i(t)- \dot\gamma(t) x_{i,2}(t) \hat\xx_1]
 	=  -\sum_{j (\neq i)} \nabla v (\xx_i(t)-\xx_j(t))  + \bm{\x}_i(t)  \ ,
 \\
 &
 	\moy{\x_{i,\m}(t)}_{\bm\xi} =0 , \qquad
 	\moy{\x_{i,\m}(t) \x_{j,\n}(s)}_{\bm\xi}
		= \delta_{ij}  \delta_{\m\n} [2 T\z  \delta(t-s) + \G_C(t,s)]
 \ .
\end{split}
\eeq
Note that the single-particle contributions associated to a finite mass and a retarded friction kernel can be straightforwardly reinstated in our effective dynamics, by a direct comparison to the summary section in Ref.~\cite{AMZ19}.

Finally, in order to study the dynamics, we need to assume that at time ${t=0}$ the particles start from a given initial configuration.
If we do not neglect inertia (${m \neq 0}$), we similarly need to specify a given distribution of initial velocities.
Note that the bare brackets ${\langle \dots \rangle}$ correspond in the whole paper to the statistical average over both the noise and this initial condition, \textit{i.e.}~to the dynamical average over trajectories of the system.
In the following, we will assume that the initial condition is sampled with a distribution isotropic \emph{outside of the shear plane}, but no such assumption is needed in the shear plane.
For explicit computations, however, one can consider the particular case of equilibrium (possibly supercooled) liquid phase at a temperature ${T_0=\beta_0^{-1}}$, where the distribution of particles is known and of high physical relevance.
The initial positions are then sampled from a Boltzmann distribution ${\propto e^{-\beta_0 V(\ul{X}(0))}}$ and the velocities from a Maxwell distribution at temperature $T_0$.
Note that we then assume implicitly that ${T_0 > T_K}$ where $T_K$ is the Kauzmann temperature below which the equilibrium liquid does not exist\footnote{Below the Kauzmann temperature one must specify which thermodynamic state is chosen~\cite{FP95,BFP97}, which complicates further the analysis.} (possibly ${T_K=0}$), and that this distribution is fully isotropic.

%-----------------
\subsection{Co-shearing frame and drifted initial condition}
\label{sec:settings-coshearing-frame}

In absence of a fluid velocity, a specific feature of the high-dimensional limit is that for finite times  particles stay `close' to their initial positions, in the sense that their relative displacements ${\xx_i(t)-\xx_i(0) \sim \mathcal{O}(1/d)}$ can be treated perturbatively in the limit ${d \to \infty}$~\cite{AMZ19}.
Physically, this can be understood from the simpler case of an isotropic random walk: each particle has so many directions towards which it can move, that it effectively explores a volume whose typical radius shrinks with an increasing dimensionality.
Note that this does not prevent the system to exhibit diffusive behaviour both in and out of equilibrium~\cite{MKZ16,KMZ16,AMZ19}.

Under a shear strain (and more generally if there is a finite fluid velocity) this argument has to be adapted because a finite shear rate can bring each particle arbitrarily far from its initial position.
We should first define the affine deformation gradient ${\hat{S}_{\gamma}(t)}$ and the corresponding `drifted initial condition' ${\RR_i'(t)}$, from which the particles will stay `close' at any given time.
In other words, we shall turn to the \emph{co-shearing frame}, in which the relative displacements ${\uu_{i}(t)=\xx_i(t)- \RR_i'(t) \sim \mathcal{O}(1/d)}$ corresponding to non-affine motion remain indeed small in high dimensions and at any finite time.
We first define the following quantities in the laboratory frame:
\beq
 \RR_i := \xx_i(0)
 \, , \quad
 \rr_{ij}(t) := \xx_i(t) - \xx_j(t)
 \, , \quad
 \rr_{0,ij} :=\rr_{ij}(0)= \RR_i - \RR_j
 \, ,
\eeq
and those relevant under shear strain:
\begin{eqnarray}
\label{eq-def-coshearing-1}
 \text{fluid velocity for the shear case:} \quad
 	&& \bm v_f(\xx,t) = \dot{\hat{\gamma}}(t) \, \xx = \dot\gamma(t) x_2 \hat{\xx}_1 \, ,
 \\
\label{eq-def-coshearing-2}
 \text{accumulated strain:} \quad
 	&& \hat{\gamma}(t) = \int_0^t \de s \, \dot{\gamma}(s) \, \hat{\xx}_1 \hat{\xx}_2^T=\gamma(t)\hat{\xx}_1 \hat{\xx}_2^T \quad \Rightarrow \quad \hat\gamma(0)= \mathbf{0} \, ,
 \\
\label{eq-def-coshearing-3}
 \text{affine deformation gradient:} \quad
 	&& \hat{S}_{\gamma}(t) = \hat{1} + \hat{\gamma}(t) \, ,
 \\
\label{eq-def-coshearing-4}
 \text{affine drifted initial positions and distances:} \quad
 	&& \RR_i'(t) = \hat{S}_{\gamma}(t) \, \RR_i \, , \quad \rr_{0,ij}'(t)= \hat{S}_{\gamma}(t) \, \rr_{0,ij} \, ,
 \\
\label{eq-def-coshearing-5}
 \text{`co-shearing frame' relative positions:}
 	&& \uu_i(t) = \xx_i(t) - \RR_i'(t) \, ,
 \\
\label{eq-def-coshearing-7}
 \text{`co-shearing frame' relative distances:}
 	&& \vv_{ij}(t) = \uu_i(t)-\uu_j(t) = \rr_{ij}(t) - \rr_{0,ij}'(t) \, ,
 \\
\label{eq-def-coshearing-9}
 \text{projection on the drifted initial distance:}
 	&& y_{ij}(t) = \frac{d}{\ell} \frac{\rr_{0,ij}'(t)}{\vert \rr_{0,ij}'(t) \vert} \cdot \vv_{ij}(t) \, .
\end{eqnarray}
These notations will become lighter when we will drop the specific indices ${(i,j)}$ in the effective stochastic processes.

This allows us to rewrite the dynamics given in Eq.~\eqref{eqC3:GENLang} as follows:
\beq
 \label{eqC3:GENLang-for-ucs}
 \begin{split}
  m \ddot \uu_{i}(t) +  \z \dot \uu_{i}(t)
 	+ \int_0^t \!\! \de s \, \G_R(t,s) \,  \dot\uu_{i}(s)
  = & - m \ddot{\hat{\gamma}}(t) \RR_i + \zeta \dot{\hat{\gamma}}(t) \uu_{i}(t) + \int_0^t \de s \, \Gamma_R(t,s) \, \dot{\hat{\gamma}}(s) \uu_{i}(s)
  \\
  	& -\sum_{j (\neq i)} \nabla v (\hat{S}_\gamma(t)(\RR_i - \RR_j) +\uu_{i}(t) - \uu_{j}(t))  + \bm{\xi}_i(t) \ ,
 \end{split}
\eeq
with a uniform initial condition ${\uu_i(0)=\mathbf{0} \, \forall i}$ and ${\arga{\RR_i}}$ the initial configuration.
If we keep the inertia finite, we should also specify the initial velocities $\arga{\dot\uu_i(0)=\dot\xx_i(0)}$.
In other words, in this co-shearing frame the fluid motion translates into
\textit{(i)}~a non-inertial force from the initial condition $\bm\RR_i$,
\textit{(ii)}~a term ${\propto \uu_{i}}$ which is an effect of the strain on the frictional force,
and \textit{(iii)}~a drift of the initial condition within the pairwise interaction ${v(|\rr_{ij}(t)|)}$.
We exploited the fact that the shear only couples two orthogonal directions, so that ${\dot{\hat{\gamma}}(s) \hat{\gamma}(s) \, \xx =\mathbf{0}}$, which cancels some of the contributions related to the initial condition.
For simplicity we set from now on ${m=0}$ and ${\Gamma_R=0}$ and consider the overdamped dynamics without retarded friction:
\beq
 \label{eqC3:GENLang-for-ucs-bis}
 \z \dot \uu_{i}(t)
   = \zeta \dot{\hat{\gamma}}(t) \uu_{i}(t) 
		-\sum_{j (\neq i)} \nabla v(\rr_{0,ij}'(t) + \uu_{i}(t) - \uu_{j}(t))
		+ \bm{\xi}_i(t)
 \, ,
\eeq
but both contributions can be reinstated directly in the effective mean-field dynamics, exactly as it was done in the companion paper~\cite{AMZ19}.
From the latter equation one readily sees that $\g=\OO(d^0)$ is a correct scaling if we expect all velocity components $u_{i,\mu}$ to scale identically with the dimension.

Our goal is to study the dynamics of Eq.~\eqref{eqC3:GENLang-for-ucs-bis} (and more generally of Eq.~\eqref{eqC3:GENLang-for-ucs}) taking first the thermodynamic limit (${N \to \infty}$) and secondly the infinite-dimensional limit (${d \to \infty}$).
Solving the dynamics consists in being able to compute dynamical observables and their time dependence, such as the average potential energy and stress tensor, the pressure, the shear stress, or the correlation and response functions~\cite{AMZ19}.
In the limit ${d \to \infty}$ the dynamics becomes exactly mean-field, and as we will show it is fully controlled by a single one-dimensional effective stochastic process, for \eg~the fluctuating quantity ${y(t)}$ defined in Eq.~\eqref{eq-def-coshearing-9}.
The novelty compared to Ref.~\cite{AMZ19} is that an external shear strain renders the dynamics anisotropic in the shear plane.
Our main message is that it can nevertheless be simply addressed once in the co-shearing frame, essentially because of the isotropy in the ${(d-2)}$ other directions.

%_____________________________________________________________
\section{Vectorial formulation of the high-dimensional mean-field dynamics}
\label{sec:vectorial-formulation}

For the isotropic case, we presented in Ref.~\cite{AMZ19} two complementary derivations in the limit ${d \to \infty}$,
via a dynamical `cavity' method (inspired by Refs.~\cite{Sz17,ABUZ18}) and a Martin-Siggia-Rose-De Dominicis-Janssen (MSRDDJ) path-integral approach (extending the equilibrium results of Refs.~\cite{MKZ16,KMZ16}).
The former is more intuitive but relies on a few unjustified assumptions, whereas the latter is technically more involved but also fully justified.
The cavity derivation can actually be straightforwardly transposed on Eqs.~\eqref{eqC3:GENLang-for-ucs}-\eqref{eqC3:GENLang-for-ucs-bis}, assuming  that ${\uu_{i}(t) \sim \mathcal{O}(1/d)}$ (Sec.~\ref{sec:vectorial-formulation-cavity}).
The physical ingredients will be the same as in Ref.~\cite{AMZ19}: \emph{in high dimension, particles stay close to ${\bm\RR_i'(t)}$, and they have many uncorrelated neighbours}.
Combined, these two ingredients allow one to build a perturbative expansion at small ${\uu_{i}(t)}$ and to invoke the central limit theorem in order to assume that some specific fluctuations are Gaussian.

From there, we obtain two vectorial effective stochastic process \emph{within the co-shearing frame}, respectively for the individual displacements of particles~${\uu(t)}$ and for the inter-particle distances~${\vv(t)}$ (Sec.~\ref{sec:vectorial-formulation-effective-stoch-processes}).
We moreover have to question the isotropy assumption of Ref.~\cite{AMZ19}, \textit{i.e.}~whether the three kernels encoding the interaction in the ${d\to\infty}$ mean-field description are diagonal at lowest order of the perturbative expansion (Sec.~\ref{sec:vectorial-formulation-breaking-stat-isotropy}).
Finally we write the coupled dynamical equations for the correlation and response functions (Sec.~\ref{sec-dynamical-equations-vectorial}).

Note that we shall only show the dynamical `cavity' method for the derivation, and briefly comment on the path-integral approach under shear strain in the conclusion.

%-----------------
\subsection{Cavity-like perturbative expansion}
\label{sec:vectorial-formulation-cavity}

We refer the reader to Ref.~\cite{AMZ19} for a thorough derivation, and provide thereafter a succinct version of it.
We start from the (simplified) dynamics Eq.~\eqref{eqC3:GENLang-for-ucs-bis}
and Taylor-expand the interaction force at lowest order in ${\uu_{i}(t)}$:
\beq
\label{eq:forceexpandcavity}
\begin{split}
 \bm{F}_{i}(t)
 =& -\sum_{j (\neq i)} \nabla v(\rr_{0,ij}'(t) + \uu_{i}(t) - \uu_{j}(t))
 \\
 =& -\sum_{j (\neq i)} \nabla v(\rr_{0,ij}'(t) - \uu_{j}(t))
 	-\sum_{j (\neq i)} \nabla \nabla^T v(\rr_{0,ij}'(t) - \uu_{j}(t)) \, \uu_i(t)
 	+ \mathcal{O}(\uu_i^2)
 \\
 =& -\sum_{j (\neq i)} \nabla v(\rr_{0,ij}'(t) - \uu_{j}(t))\vert_{\uu_i=0}
 	-\int_0^t \de s \sum_{j (\neq i)} \frac{\delta \nabla v(\rr_{0,ij}'(t) - \uu_{j}(t))}{\delta \uu_i(s)}\Big\vert_{\uu_i=0} \uu_i(s)
 \\ &	-\sum_{j (\neq i)} \nabla \nabla^T v(\rr_{0,ij}'(t) - \uu_{j}(t)) \Big\vert_{\uu_i=0} \, \uu_i(t)
 	+ \mathcal{O}(\uu_i^2)
 \, .
\end{split}
\eeq
The second line is obtained by fixing the particle $i$ at its drifted initial position ${\RR_i'(t)}$ and treating its displacement ${\uu_i(t)}$ as an external field.
However, the full trajectories of the other particles still depend implicitly on ${\lbrace \uu_i(s) \rbrace_{s \in [0,t]}}$ via the coupled Langevin dynamics, hence the additional Taylor expansion in the third line~\cite{AMZ19}.
The notation ${\dots \vert_{\uu_i=0}}$ corresponds physically to fixing the particle ${i}$ at its drifted initial position at all times, \textit{i.e.}~${\xx_i(s)=\RR_i'(s) \, \forall s \in [0,t]}$.
Renaming the different contributions and truncating the expansion at order ${\mathcal{O}(\uu_i)}$, we have then
\beq
\begin{split}
 & \bm{F}_{i}(t)
 %= -\sum_{j (\neq i)} \nabla_{\uu_{i}(t)} v(\uu_{i}(t) - \uu_{j}(t) + \rr_{0,ij}'(t))
 \approx 	\tilde{\bm{F}}_i^f(t)
 			+ \int_0^t \de s \, \hat{M}_{R,i}(t,s) \, \uu_{i}(s)
 			- \hat{k}_i(t) \, \uu_{i}(t)
 \, ,
 \\
 & \text{with} \quad
	\moy{\tilde{\bm{F}}_i^f(t)} \equiv \sqrt2\,{\bm{\bar\Xi}}_i(t) \, , \quad
	\moy{\tilde{F}_{i \mu}^f(t)\tilde{F}_{i \nu}^f(s)} \equiv M_{C,i}^{\mu\nu}(t,s) \, .
\end{split}
\eeq
Physically,
${\tilde{\bm{F}}_i^f(t)}$ is the force resulting of the random kicks of all the other particles on particle~$i$ if it were not moving from ${\RR_i'(t)}$,
with the vector ${{\bm{\bar\Xi}}_i(t)}$ the resulting overall drift and the memory kernel ${\hat{M}_{C,i}(t,s)}$ encoding its correlation in time.
The response memory kernel ${\hat{M}_{R,i}(t,s)}$ characterises the response at time~$t$ of all the other particles with respect to an infinitesimal displacement of particle~$i$ at an intermediate time~${s\in[0,t]}$.
${\hat{k}_i(t)}$ corresponds to the spring constant (tensorial here) of the restoring force -- originating from the surrounding particles -- felt by the fixed particle~$i$ if it moves weakly at time~$t$.
We emphasise that this description is expressed in the co-shearing frame, so that the actual positions of particles have to be reconstructed using ${\xx_i(t) = \uu_{i}(t) + \RR_i'(t) = \uu_{i}(t) + \argp{\hat{1} + \hat{\gamma}(t)} \RR_i }$.

Because each particle has many neighbours (of order $d$), by the central limit theorem the distributions of the three kernels ${\lbrace \hat{k}_i(t),\hat{M}_{C,i}(t,s),\hat{M}_{R,i}(t,s) \rbrace}$ are peaked at their mean values, rendering the mean-field description exact for ${d\to\infty}$.
These mean values are in fact defined as dynamical averages with the particle~$i$ fixed, by construction of the Taylor expansion.
However, in a mean-field description we can simply remove this constraint when averaging over all the pairs of $N$ particles and use the actual dynamical average ${\moy{\bullet}}$ over the possible trajectories~\cite{AMZ19}:
\beq
\begin{split}
 k_i^{\mu \nu}(t)
 	& \approx \sum_{j (\neq i)} \moy{\nabla_\mu \nabla_\nu v(\rr_{0,ij}'(t) - \uu_{j}(t))}_{i \text{ fixed}}
 	\approx \frac{1}{N} \sum_{i \neq j} \moy{\nabla_\mu \nabla_\nu v (\rr_{ij}(t))}
 	\equiv k^{\mu\nu}(t)
 \, ,
 \\
 M_{C,i}^{\mu\nu}(t,s)
 	& \approx \sum_{j (\neq i)} \moy{\nabla v(\rr_{0,ij}'(t) - \uu_{j}(t))\nabla v(\rr_{0,ij}'(s) - \uu_{j}(s))}_{i \text{ fixed}}
 	\approx \frac{1}{N} \sum_{j \neq i} \moy{ \nabla v(\rr_{ij}(t)) \nabla v (\rr_{ij}(s)}
 	\equiv M_C^{\mu\nu}(t,s)
 	\, ,
 \\
 M_{R,i}^{\mu\nu}(t,s)
 	& \approx- \sum_{j (\neq i)} \frac{\moy{ \delta \nabla_\mu v(\rr_{0,ij}'(t) - \uu_{j}(t))}_{i \text{ fixed}}}{\delta u_{i,\nu}(s)} \Big\vert_{\uu_i=0}
 	\approx \frac{1}{N} \sum_{i \neq j} \left. \frac{\d \moy{  \nabla_\m v(\rr_{ij}(t)) }_{\bm{P}}}{\d P_{\nu,ij}(s)}\right\vert_{\bm P=\bm0}
 	\equiv M_{R}^{\mu\nu}(t,s)
 	\, .
\end{split}
\eeq
In the definition of $\hat{M}_R$ the term ${\bm{P}_{ij}(s)}$ is added in the dynamics as a perturbation within the interaction potential as ${\nabla v(\rr_0'(t)+\vv_{ij}(t)- \bm{P}_{ij}(t))}$, as denoted by the corresponding dynamical average ${\moy{\bullet}_{\bm{P}}}$.
By the central limit theorem the random forces ${\tilde{\bm{F}}_i^f(t)}$ moreover have a Gaussian statistics, of two-time correlation ${\hat{M}_C(t,s)}$ and mean:
\beq
 \moy{\tilde{\bm{F}}_i^f(t)}
 	=-\sum_{j (\neq i)} \moy{\nabla v(\rr_{0,ij}'(t) - \uu_{j}(t))\vert_{\uu_i=0}}_{i \text{ fixed}}
 	\approx - \frac{1}{N} \sum_{j \neq i} \moy{\nabla v(\rr_{ij}(t))}
 	\equiv \sqrt{2}\,{\bm{\bar\Xi}}(t)
 \, .
\eeq
Note that there could be an overall drift in the shear plane, which is straightforwardly zero in the isotropic case discussed in the companion paper~\cite{AMZ19}.
Thereafter we choose to subtract ${\bar{\bm\Xi}(t)}$ from the definition of the noise, so the latter is of zero mean.

We can thus rewrite Eq.~\eqref{eqC3:GENLang-for-ucs-bis}, at lowest order in the fluctuations $\uu$, as:
\beq
 \label{eqC3:GENLang-for-ucs-ter}
 \begin{split}
 & \z \dot \uu_{i}(t)
	= 	- (\hat{k}(t)-\zeta \hat{\dot{\gamma}}(t)) \uu_{i}(t)
 		+ \int_0^t \de s \, \hat{M}_R(t,s) \, \uu_{i}(s)
                + \sqrt{2}\,{\bm{\bar\Xi}}(t)
 		+ \sqrt{2}\,\bm{\Xi}_i(t)
 \, , \\
 &  \text{with} \quad \sqrt{2}\,\bm{\Xi}_i(t) = \tilde{\bm{F}}_i^f(t) + \bm{\x}_i(t)-\sqrt{2}\,{\bm{\bar\Xi}}(t)
 \, .
 \end{split}
\eeq
For the relative inter-particle distances in the co-shearing frame ${\vv_{ij}(t)=\uu_i(t)-\uu_j(t)}$, one can perform a similar derivation although one has to keep an extra potential contribution at this order of the high-dimensional expansion\footnote{In high $d$ this contribution is subdominant but needed to get the correct dynamics of the projected motion defined in Eq.~\eqref{eq-def-coshearing-9}, see Sec.~\ref{sec:scalar-formulation-eff-Langevin-process-longitud}; see also Ref.~\cite[footnote 8]{AMZ19}.}:
\beq
 \label{eqC3:GENLang-for-wcs-ter}
 \begin{split}
 & \frac{\z}{2} \dot \vv_{ij}(t)
	= 	- \frac12 (\hat{k}(t)-\zeta \hat{\dot{\gamma}}(t)) \vv_{i}(t)
 		+ \frac12 \int_0^t \de s \, \hat{M}_R(t,s) \, \vv_{i}(s)
 		- \nabla v(\rr_{0,ij}'(t) + \vv_{ij}(t))
 		+  \bm{\Xi}_{ij}(t)
 \, , \\
 & \text{with} \quad \bm{\Xi}_{ij}(t) = [\bm{\Xi}_i(t)-\bm{\Xi}_j(t)]/\sqrt2
 \, .
 \end{split}
\eeq
Assuming uncorrelated neighbours, the noise ${{\bm\Xi}_{ij}(t)}$ has the same Gaussian statistics as the individual noises ${\bm\Xi}_i(t)$ and ${\bm\Xi}_j(t)$, hence removing the specific indices from now on we have:
\beq
 \moy{{\bm \Xi}(t)}_{\bm\Xi}=\mathbf{0}
 \, , \quad
 \moy{\Xi_\mu(t) \Xi_\nu(s)}_{\bm\Xi}= \delta_{\mu\nu} \argc{T \zeta \delta(t-s) + \frac12 \Gamma_C(t,s)} + \frac12 M_C^{\mu\nu}(t,s)-{\bar\Xi}_\m(t){\bar\Xi}_\n(s)
 \, .
\eeq
The stochastic processes being the same for all particles, in the mean-field description we can remove the indices for ${\uu_i(t)}$ and ${\vv_{ij}(t)}$ as well.

%-----------------
\subsection{Effective stochastic processes}
\label{sec:vectorial-formulation-effective-stoch-processes}

The cavity-like derivation of the previous section can thus be summarised in the following set of effective stochastic processes in the co-shearing frame, valid in the high-dimensional limit, with the accumulated strain ${\hat{\gamma}(t) = \int_0^t \de s \,  \dot{\gamma}(s) \, \hat{\xx}_1 \hat{\xx}_2^T}$ in the laboratory frame and the affine deformation of the initial condition ${\rr_0'(t) = (\hat{1} + \hat{\gamma}(t))\rr_0}$:
\beq
\label{eq-coshearing-effective-stoch-processes}
\begin{split}
  & \z \dot \uu(t)
  = - (\hat{k}(t)-\zeta \hat{\dot{\gamma}}(t)) \uu(t)
 	+ \int_0^t \de s \, \hat{M}_R(t,s) \, \uu(s)
 	+ \sqrt 2\,\bm{\bar \Xi}(t)
 	+ \sqrt{2}\,\bm{\Xi}(t)
 \, ,
 \\
  & \frac{\z}{2} \dot \vv(t)
  =	- \frac12 (\hat{k}(t)-\zeta \hat{\dot{\gamma}}(t)) \vv(t)
 	+ \frac12 \int_0^t \de s \, \hat{M}_R(t,s) \, \vv(s)
 	- \nabla v \argp{\rr_{0}'(t) + \vv(t)}
 	+ \bm{\Xi}(t)
 \, ,
 \\
 & \uu(0)=0 \, , \quad \vv(0)=0 \, ,
 \\
 &	\moy{\X_{\m}(t)}_{\bm{\Xi}}=0 \, , \quad
	\moy{\X_{\m}(t) \X_{\n}(s)}_{\bm{\Xi}}
	= \d_{\m\n} \left[ T \z \d(t-s) +\frac12 \G_C(t,s)\right] +\frac12 M^{\m\n}_C(t,s)-{\bar\Xi}_\m(t){\bar\Xi}_\n(s)
 \, ,
\end{split}
\eeq
with the following self-consistent equations for the kernels and the overall drift:
\beq
\label{eqC3:Mself}
\begin{split}
 k^{\m\n}(t)
 	&= \r \int \de\rr_0 \, g_{\text{in}}(\rr_0) \, \moy{ \nabla_\m \nabla_\n v( \rr_0'(t) + \vv(t))}_{\vv}
 \ , \\
 M^{\m\n}_C(t,s)
 	&=  \r \int \de \rr_0 \, g_{\text{in}}(\rr_0) \, \moy{ \nabla_\m v(\rr_0'(t) + \vv(t))  \nabla_\n v(\rr_0'(s) + \vv(s)) }_{\vv}  
 \ , \\
 M^{\m\n}_R(t,s)
 &= \r  \int \de \rr_0 \, g_{\text{in}}(\rr_0) \, \left. \frac{\d \moy{  \nabla_\m v(\rr_0'(t) + \vv(t)) }_{\vv,\bm{P}}}{\d P_{\nu}(s)}\right\vert_{\bm P=\bm0}
 \, , \\
 \sqrt2\,{\bm{\bar\Xi}}(t)
 &= - \r  \int \de \rr_0 \, g_{\text{in}}(\rr_0) \, \moy{\nabla v(\vert \rr_0'(t) + \vv(t) \vert)}_{\vv}
 \ .
\end{split}
\eeq
Here $\rho$ is the density and ${g_{\text{in}}(\rr_0)}$ the distribution of inter-particle distances in the laboratory frame at initial time ${t=0}$~\cite{AMZ19,hansen}.
For the particular case of an initial equilibrium at inverse temperature ${\Bi}$ one has ${g_{\text{in}}(\rr_0)=g_{\rm{eq}}(\rr_0)=e^{-\Bi v( \rr_0 )}}$~\cite{FRW85,WRF87,FP99}.
The brackets ${\moy{\bullet}_{\vv}}$ denote the dynamical average over the stochastic process $\vv(t)$ (thus self-consistently defined), from a given initial condition $\rr_0$.
${\bm{P}(s)}$ is added in the dynamics as a perturbation field inside the interaction potential ${\nabla v(\rr_0'(t)+\vv(t) - \bm{P}(t))}$.

We emphasise that this type of `dynamical cavity' derivation is quite generic, 
since it is essentially a Taylor expansion of the many-body Langevin dynamics, in a self-consistent mean-field formulation~\cite{MPV87,ABUZ18}.
As in Ref.~\cite{AMZ19}, it thus relies on identifying the `small field' for the expansion, and arguing that this mean-field description is exact because of a central limit theorem applies to the (\textit{a priori} fluctuating) kernels and parameters of the expansion.

%-----------------
\subsection{Breaking of the statistical isotropy in the shear plane}
\label{sec:vectorial-formulation-breaking-stat-isotropy}

If both the dynamics and initial condition were isotropic, the mean-field kernels ${\lbrace \hat{k}(t),\hat{M}_{C}(t,s),\hat{M}_{R}(t,s) \rbrace}$ would be simply diagonal, as discussed in the companion paper~\cite{AMZ19}.
Under a shear strain ${\dot{\hat{\gamma}}(t) \, \xx = \dot{\gamma}(t) \, \xx_2 \, \hat{\xx}_1}$, the rotational symmetry of Eqs.~\eqref{eq-coshearing-effective-stoch-processes}-\eqref{eqC3:Mself} is broken exclusively in the two-dimensional shear plane, hence the kernels are diagonal except in the sector ${\lbrace \hat\xx_1,\hat\xx_2 \rbrace}$.
They have thus the following structure:
\beq
 \hat{k}(t)
 	=	\left\lbrace \begin{array}{cccccc}
 		k^{11}(t) & k^{12}(t) & 0 & 0 & \cdots & 0
 		\\
 		k^{21}(t) & k^{22}(t) & 0 & 0 & \cdots & 0
 		\\
 		0 & 0 & k^{\text{iso}}(t) & 0 & \cdots & 0
 		\\
 		\vdots & \vdots & 0 & \ddots & \ddots & \vdots
 		\\
 		\vdots & \vdots & \vdots & \ddots & \ddots & 0
 		\\
 		0 & 0 & 0 & \cdots & 0 & k^{\text{iso}}(t)
 		\end{array} \right\rbrace
 \quad \Leftrightarrow \quad
 \hat{k}(t) := \arga{k^{\text{iso}}(t),k^{11}(t),k^{12}(t),k^{21}(t),k^{22}(t)}
 \, ,
\eeq
and similarly for the memory kernels ${\hat{M}_{C,R}(t,s):= \arga{M_{C,R}^{\text{iso}}(t),M_{C,R}^{11}(t),M_{C,R}^{12}(t),M_{C,R}^{21}(t),M_{C,R}^{22}(t)}}$.
The overall drift ${{\bm{\bar\Xi}}(t)}$ has \textit{a priori} two non-zero components ${\lbrace \bar{\Xi}_1(t), \bar{\Xi}_2(t) \rbrace}$ (that vanish in the isotropic case).

The kernels and drift definitions are all related to the forces between particles, so we need to examine statistical averages of simple and double derivatives of the interaction potential in high dimension.
We use that ${\mathcal{O}(\vv) = \mathcal{O}(\uu) = \mathcal{O}(1/d)}$ to rewrite
first for the unitary inter-particle distance: 
\beq
\label{eq-r0-r0t-unitary-vectors-bis-bis}
 \hat{\rr}(t)
 = \frac{\rr(t)}{\vert \rr(t) \vert}
 = \frac{\rr_{0}'(t)+\vv(t)}{\vert \rr_{0}'(t)+\vv(t) \vert}
 \approx \hat{\rr}_0'(t)
 \, .
\eeq
Since ${\hat \rr_0^2=1}$, its correct scaling is ${\hat r_{0,\m}=g_\m/\sqrt{d}}$ where ${g_\m=\OO(1)}$.
Then for the drifted initial condition:
\beq
\label{eq-r0-r0t-unitary-vectors-bis}
\begin{split}
 & \rr_0'(t)
 \stackrel{\eqref{eq-def-coshearing-4}}{=} \argp{\hat{1} + \hat{\gamma}(t)} \rr_0
 	= \rr_0 + \gamma(t) r_{0,2} \hat{\xx}_1
 \\
 & \Rightarrow \quad r_0'(t)
 = \valabs{\rr_0'(t)}
 = r_0 \sqrt{1 + 2 \gamma(t) \frac{r_{0,1} r_{0,2}}{r_0^2} + \gamma(t)^2 \frac{r_{0,2}^2}{r_0^2}}
 \approx r_0 \argp{1 + \frac{\gamma(t)}{d} \, g_1 g_2 + \frac{\gamma(t)^2}{2d} \, g_2^2}
 \\
 & \Rightarrow \quad \hat{\rr}_0'(t)
 = \frac{{\rr}_0'(t)}{\valabs{{\rr}_0'(t)}}
 \approx \frac{\argp{\hat{1} + \hat{\gamma}(t)} \rr_0}{r_0 \argp{1 + \mathcal{O}(1/d)}}
 \approx  \argp{\hat{1} + \hat{\gamma}(t)} \hat{\rr}_0
%  \approx \frac{1}{\sqrt{d}} \argp{g_\mu + \gamma(t)g_2 \, \delta_{1,\mu}} \hat{\xx}_\mu
 \, .
\end{split}
\eeq
The normalisation of the above unitary vector is correct only in the infinite-dimensional limit.
Using that the interaction potential is radial\footnote{We recall the following identities for the derivatives of a radial function:
${\nabla v \argp{\valabs{\rr}}
 	= v'\argp{\valabs{\rr}} \hat{\rr}
 	\, , \quad 
 \nabla_{\mu} \nabla_{\nu} v\argp{\valabs{\rr}}
 	= v''\argp{\valabs{\rr}} \hat{r}_\mu \hat{r}_\nu + v'\argp{\valabs{\rr}} \frac{\delta_{\mu\nu} - \hat{r}_\mu \hat{r}_\nu}{\valabs{\rr}}}$
 .
}
we thus obtain in high dimension\footnote{The potential $v(\valabs{\rr_0'(t) + \vv(t)})$ must not be approximated further, because it changes by a finite quantity when its
argument changes by $\OO(1/d)$, as expressed by Eq.~\eqref{eq:potscal}. See the companion paper~\cite{AMZ19} for a more detailed discussion.}:
\beq
\label{eq-derivatives-radial-functions-refs}
\begin{split}
 \nabla_\mu v \argp{\valabs{\rr(t)}}
 	& \approx  v'\argp{\valabs{\rr_0'(t) + \vv(t)}} \hat\rr_0'(t)
 	\, ,
\\
 \nabla_\mu \nabla_\nu v \argp{\valabs{\rr(t)}}
 	& \approx v''\argp{\valabs{\rr_0'(t)+\vv(t)}} \hat{r}_{0,\mu}'(t) \hat{r}_{0,\nu}'(t) + \delta_{\mu\nu} \frac{v'\argp{\valabs{\rr_0'(t)+\vv(t)}}}{\valabs{\rr_0'(t)+\vv(t)}}
 \, .
\end{split}
\eeq
All the kernels and drift coefficients can then be reconstructed from statistical averages --~over the stochastic process for~$\vv(t)$ and over the initial condition~$\rr_0$~-- of combinations of these two quantities.
The corresponding explicit expressions are given in Appendix~\ref{sec-appendix-explicit-coeff}, and they will be further simplified in Sec.~\ref{sec-scalar-formulation} in the infinite-dimensional limit.

Under these symmetry considerations, we can rewrite the effective stochastic process~\eqref{eq-coshearing-effective-stoch-processes} for the individual relative displacements as
\beq
\label{eq-coshearing-effective-stoch-processes-bis-bis}
\begin{split}
 & \left\lbrace \begin{array}{l}
  \z \dot u_{1}(t)
  = - k^{11}(t) \, u_{1}(t)
  	- \underbrace{\argp{ k^{12}(t) - \zeta \dot{\gamma}(t)} \, u_{2}(t)}
 	+ \int_0^t \de s \, \argc{ {M}^{11}_R(t,s) \, u_{1}(s) + \underbrace{{M}^{12}_R(t,s) \, u_{2}(s)}}
 	+\overbrace{ \sqrt{2}\, \bar\Xi_1(t)+ \sqrt{2} \,\Xi_1(t)}
 \\
  \z \dot u_{2}(t)
  = - k^{22}(t) \, u_{2}(t)
  	- \underbrace{k^{21}(t) \, u_{1}(t)}
 	+ \int_0^t \de s \, \argc{ {M}^{22}_R(t,s) \, u_{2}(s) + \underbrace{{M}^{21}_R(t,s) \, u_{1}(s)}}
 	+ \overbrace{ \sqrt{2}\, \bar\Xi_2(t)+ \sqrt{2} \,\Xi_2(t)}
 \\ \\
  \z \dot u_{\mu}(t)
  = - k^{\text{iso}}(t) \, u_{\mu}(t)
 	+ \int_0^t \de s \, {M}^{\text{iso}}_R(t,s) \, u_{\mu}(s)
 	+ \sqrt{2} \,\Xi_\mu(t)
 	\, , \quad \text{for }  \mu \in \arga{ 3, \dots, d} \, .
 \end{array} \right.
\end{split}
\eeq
The shear strain in the plane ${\lbrace \hat\xx_1, \hat\xx_2 \rbrace}$ introduces additional couplings (underbraced terms) as well as a drift and a renormalised noise (overbraced).
Out of the shear plane, the process is formally the same as in the isotropic case~\cite[Eq.(24)]{AMZ19}, except that here the `isotropic' kernels do depend upon the shear strain.
Similarly for the effective stochastic process for $\vv(t)$:
\beq
\label{eq-coshearing-effective-stoch-processes-bis-bis-w}
 \z \dot \vv(t)
  = - (\hat{k}(t)-\zeta \dot{\hat{\gamma}}(t)) \vv(t)
 	+ \int_0^t \de s \, \hat{M}_R(t,s) \, \vv(s)
 	- v'(\vert \rr_0'(t)+\vv(t )\vert) \, \argp{\hat\rr_0 + \gamma(t) \hat{r}_{0,2} \hat\xx_1}
 	+ \bm{\Xi}(t)
 \, .
\eeq

As such, the mean-field vectorial equations~\eqref{eq-coshearing-effective-stoch-processes-bis-bis}-\eqref{eq-coshearing-effective-stoch-processes-bis-bis-w} (together with the kernels given in Appendix~\ref{sec-appendix-explicit-coeff}) are quite generic: they assume small displacements in the co-shearing frame, independent neighbours and an isotropic initial condition outside of the shear plane.
They are closed self-consistently through a central limit theorem which strictly applies with a diverging number of neighbours (\ie first ${N \to \infty}$ and secondly ${d\to\io}$).

%-----------------
\subsection{Dynamical equations for the matricial correlation and response functions}
\label{sec-dynamical-equations-vectorial}

To close the dynamics, we now write the infinite-dimensional dynamical equations for the correlation and response functions ${\lbrace \hat{C},\hat{R} \rbrace}$ in the co-shearing frame.
They are matrices if no isotropy assumption is made, and are defined as follows:
\beq
\label{eq-def-correlation-response-functions-matricial}
\begin{split}
 	C^{\mu\nu}(t,t')
 	&= \frac{1}{N} \sum_{i=1}^N \moy{u_{i,\mu}(t) u_{i,\nu} (t')}
 \, ,
 \\
 	R^{\mu\nu}(t,t')
 	&= \frac{1}{N} \sum_{i=1}^N \frac{\delta \moy{u_{i,\mu}(t)}}{\delta \lambda_{i,\nu}(s)}\big\vert_{\arga{{\bm\lambda}={\bm 0 }}}
 \, ,
\end{split}
\eeq
with their isotropic counterparts, as defined in Ref.~\cite{AMZ19}:
\beq
\label{eq-def-correlation-response-functions-matricial-isotropic}
\begin{split}
	C^{\text{iso}}(t,t')
 	&= \frac 1d \sum_{\mu=1}^d C^{\mu\mu}(t,t')
	= \frac{1}{Nd} \sum_{i=1}^N \moy{\uu_{i}(t) \cdot \uu_{i} (t')}
 \, ,
 \\
	R^{\text{iso}}(t,t')
 	&= \frac 1d\sum_{\mu=1}^d R^{\mu\mu}(t,t')
 	= \frac{1}{Nd} \sum_{i=1}^N\sum_{\mu=1}^d \frac{\delta \moy{u_{i,\mu}(t)}}{\delta \lambda_{i,\mu}(s)}\big\vert_{\arga{{\bm\lambda}={\bm 0 }}}
 \, .
\end{split}
\eeq
The mean-square-displacement (MSD) function ${D(t,t')}$ is  defined with respect to the isotropic part of the correlation function:
\beq
\label{eq-def-MSD-functions-bis}
\begin{split}
 & \DE(t,t')
 	= \frac{1}{Nd} \sum_{i=1}^N \moy{\argc{\uu_i(t)-\uu_i(t')}^2}
 	= C^{\text{iso}}(t,t)+C^{\text{iso}}(t',t') - 2 C^{\text{iso}}(t,t')
 \, ,
 \\
 & \DE_r(t) = \DE(t,0) \, .
\end{split}
\eeq
Under shear strain, the matrices ${\hat{C}(t,t')}$ and ${\hat{R}(t,t')}$ have the same structure as the kernels ${\lbrace\hat{k}(t),\hat{M}_C(t,t'),\hat{M}_R(t,s)\rbrace}$,
so their only non-zero terms are ${\arga{C^{\mu\mu},C^{12},C^{21}}}$ and ${\arga{R^{\mu\mu},R^{12},R^{21}}}$.

In order to obtain the dynamical equations for ${\{\hat{C},\hat R\}}$, we start from the stochastic process for ${{\uu}_i}$ in Eq.~\eqref{eq-coshearing-effective-stoch-processes} and make use of the following relations for a Gaussian noise (cf.~Ref.~\cite[Eq. (95)]{AMZ19} and references therein towards an explicit proof of these relations):
\beq
\begin{split}
 R^{\mu\nu}(t,t')
	&= \frac{1}{N} \sum_{i=1}^N \moy{\frac{\delta u_{i,\mu}(t)}{\delta \argp{\sqrt{2} \,\Xi_{i,\nu}(t')}}} \ ,
 \\
 \moy{\sqrt{2}\, \Xi_{i,\mu}(t) u_{i,\nu}(t')}
 	&= \int_0^{t'} \de s \, \sum_{\nu'=1}^d \moy{\sqrt{2} \,\Xi_{i,\mu}(t) \sqrt{2}\, \Xi_{i,\nu'}(s)} R^{\nu \nu'}(t',s)
 \\
 	&= \int_0^{t'} \de s \, \sum_{\nu'=1}^d \argc{ \delta_{\mu\nu'} \argp{2T\zeta \delta(t-s) + \Gamma_C(t,s)} + M_C^{\mu\nu'}(t,s) } R^{\nu \nu'}(t',s)
 	\, .
\end{split}
\eeq
%
% As for the dynamical equation for ${R(t,t')}$, we start again from the stochastic process $\arga{\zeta \dot{\uu}_{i,\mu}(t)}_{\mu}$ and we simply take its functional derivative with respect to $\delta(\sqrt{2} \Xi_{i\nu}(t'))$ and its dynamical average $\moy{\bullet}$.
%
Similarly to Ref.~\cite[Sec. 6]{AMZ19} we then obtain:
\begin{equation}
\label{eq:CRfinite}
\begin{split}
 \z\left[ \frac{\partial }{\partial t} \hat C(t,t')- \hat {\dot{\g}}(t) \hat C(t,t') \right]
  	=&2T\z \hat R^T(t',t)-\hat k(t) \hat C(t,t')+\int_0^{t} \de s\,\hat M_R(t,s)\hat C(s,t')\\
  	&+\int_0^{t'}\de s\,\left[\G_C(t,s) \hat{1}+\hat M_C(t,s)\right]\hat R^T(t',s)
 \ ,\\
 \z\left[ \frac{\partial }{\partial t} \hat R(t,t') - \hat {\dot{\g}}(t) \hat R(t,t') \right]
 	=&\d(t-t') \hat{1} - \hat k(t) \hat R(t,t')+\int_{t'}^{t} \de s\,\hat M_R(t,s) \hat R(s,t')
 \ .
\end{split}
\end{equation}
Integration intervals are truncated owing to the causality of response functions~\cite{pottier,Cu02}.

Note that in the right-hand side of the first line describing the correlations dynamics we have discarded an additional term $\hat E(t,t')$ which has non-zero components only in the shear plane ${\{\hat\xx_1,\hat\xx_2\}}$:
\beq
E^{\m\n}(t,t')=\sqrt2\,\bar\Xi_\m(t)\moy{u_\n(t')}
\eeq
with ${\bf{\bar\Xi}}$ the drift defined in Eq.~\eqref{eqC3:Mself}, and $\moy{u_{1,2}(t')}$ is solution of the first-order linear differential two-dimensional system given by averaging the first two lines of Eq.~\eqref{eq-coshearing-effective-stoch-processes-bis-bis} over the noise ${\bf\Xi}$.
The reason is that for $d\to\io$ it is subdominant due to the scaling of ${\bf{\bar\Xi}}$, see Appendix~\ref{sec-appendix-explicit-coeff}.
We can thus write down the correlation and response evolution for the different components, by substituting the kernels coefficients obtained in Appendix~\ref{sec-appendix-explicit-coeff}.

The equations for the isotropic scalar components ${\{C^{\text{iso}},R^{\text{iso}}\}}$ are obtained from them through the definition~\eqref{eq-def-correlation-response-functions-matricial-isotropic}, summing over all diagonal components.
This operation simplifies the resulting equations with respect to Eqs.~\eqref{eq:CRfinite}: the key idea is that in the ${d\to\io}$ limit other components than the isotropic ones are subdominant due to their small number.
We have for instance:
\beq
\begin{split}
 \frac{1}{d} \sum_{\mu\nu} k^{\mu\nu}(t) C^{\nu\mu}(t,t')
 =& \left(1-\frac2d\right)k^{\text{iso}}(t) C^{\text{iso}}(t,t')
 		+ \frac{1}{d} k^{11}(t)C^{11}(t,t')+ \frac{1}{d} k^{22}(t)C^{22}(t,t')
 		+\frac2d k^{12}(t)C^{12}(t,t')\\
 		\stackrel{(d\to\io)}{=}& k^{\text{iso}}(t) C^{\text{iso}}(t,t')
 \, .
\end{split}
\eeq
We eventually get:
\beq
\label{eq:CRfinite-isotropic}
\begin{split}
\z \partial_t C^{\text{iso}}(t,t')
  	&= 2 T \zeta R^{\text{iso}}(t',t)
  			- k^{\text{iso}} (t) C^{\text{iso}}(t,t')
  			+ \int_0^{t} \de s \, M_R^{\text{iso}}(t,s) C^{\text{iso}}(s,t')
  			+ \int_0^{t'} \de s \, \argc{\Gamma_C(t,s) + M_C^{\text{iso}}(t,s)} R^{\text{iso}}(t',s)
  \, ,
  \\
\z \partial_t R^{\text{iso}}(t,t')
 	&= \delta(t-t') - k^{\text{iso}}(t) R^{\text{iso}}(t,t') + \int_{t'}^t \de s \, M_R^{\text{iso}}(t,s) R^{\text{iso}}(s,t')
 \, .
\end{split}
\eeq
These equations formally possess the exact same structure as the one obtained in absence of shear in Ref.~\cite[Eq.(96)]{AMZ19}.
However here the `isotropic' scalar kernels are affected by the shear strain through \eg~the gap process ${\rr_0'(t)+\vv(t)}$ defining them self-consistently.

We show in the next section
how the present mean-field description can be further simplified in the infinite-dimensional limit into a scalar stochastic process depending on isotropic kernels, providing a rather simple description of the shear strain protocol in this limit.
This is possible because the anisotropy in the shear plane is to be compared to the large number ${(d-2)}$ of other directions with statistical isotropy.

%_____________________________________________________________
\section{Scalar formulation of the infinite-dimensional effective dynamics}
\label{sec-scalar-formulation}

In the isotropic case~\cite{AMZ19}, one can simplify further the vectorial effective stochastic processes~\eqref{eq-coshearing-effective-stoch-processes-bis-bis}-\eqref{eq-coshearing-effective-stoch-processes-bis-bis-w} into a \emph{scalar} process, benefiting from the rotational invariance of the pairwise potential ${v(\rr)}$.
We show thereafter that the argument could be applied to the anisotropic case under shear as well, using again the statistical isotropy which prevails outside the shear plane.
The proof follows ideas initially presented in Ref.~\cite{BU18}.

We first recall the high-dimensional scaling of the relevant quantities and of the dynamical equations (Sec.~\ref{sec:highd-scaling}).
Next, we show how the gap can be simply decomposed into three additive contributions, given by the initial condition, a `longitudinal' motion, and a `transverse' diffusive motion (Sec.~\ref{sec:scalar-formulation-decomposition-gap}).
Then we are able to write down the effective Langevin process for the `longitudinal' motion (Sec.~\ref{sec:scalar-formulation-eff-Langevin-process-longitud}),
and consequently reconstruct the process for the gap itself.
We provide the self-consistent definitions of the rescaled kernels (Sec.~\ref{sec:scalar-formulation-rescaled-kernels}) and we wrap-up the physical picture we will have obtained (Sec.~\ref{sec-scalar-formulation-wrapup}).

%-----------------
\subsection{High-dimensional scaling and dynamical equations}
\label{sec:highd-scaling}

From now on, we will explicitly write that the potential is radial, ${v(\rr)=v(\vert \rr \vert)}$.
We first recall the definition of the rescaled interaction potential in high dimension given in Eq.~\eqref{eq:potscal}:
\beq
\label{eq-def-rescaled-potential-1}
 v(r) = \bar{v}(h)
 \, , \quad
 r = \ell(1+h/d)
 \, ,
\eeq
where $\ell$ is the typical interaction scale (which sets the length unit) and $h$ is the rescaled gap between two interacting particles~\cite{KMZ16,MKZ16,CKPUZ17}. 
The physics behind these definitions is that in the limit ${d\to\infty}$ typical relative distances ${\vv(t)}$ are of ${\mathcal{O}(1/d)}$, hence in this limit we need to properly rescale the potential to focus on the fluctuating gap ${h(t)\sim \mathcal{O}(1)}$.
This rescaling is mandatory to obtain a non-trivial dynamics.

As in Ref.~\cite{AMZ19} and following Refs.~\cite{KMZ16,MKZ16}, in high $d$ we define the following rescaled parameters and kernels: 
\beq
\label{eqC3:fricscal2-shear}
\begin{split}
 & \wh m = \frac{ \ell^2}{2d^2} m
 \ , \quad
 \wh\z = \frac{ \ell^2}{2d^2} \z 
 \ , \quad
 \GG_C(t,s) = \frac{ \ell^2}{2d^2} \G_C(t,s)
 \, , \quad
 \GG_R(t,s) = \frac{ \ell^2}{2d^2} \G_R(t,s)
 \, , \\
& \hat\CC(t,t')=\frac{d^2}{\ell^2}\hat C(t,t') \ ,
\qquad 
\hat \cR(t,t')=\frac{d^2}{\ell^2}\hat R(t,t') \ ,
\qquad 
 \D(t,t')=\frac{d^2}{\ell^2}\DE(t,t')
 \, ,
 \\
 & \hat\k(t) = \frac{ \ell^2}{2d^2}\hat k(t)
 \ , \quad
\hat \MM_C(t,s) = \frac{ \ell^2}{2d^2}\hat M_C(t,s)
 \ , \quad
 \hat\MM_R(t,s) = \frac{ \ell^2}{2d^2}\hat M_R(t,s)
 \ .
\end{split}
\eeq
As a consequence we retrieve the same formal structure for the dynamical equations as in Ref.~\cite[Eq.(104)]{AMZ19}:
\begin{equation}
\label{eq:scaldyneq-copy-from-AMZ19}
\begin{split}
 \wh\z\frac{\partial }{\partial t}\CC^{\text{iso}}(t,t')
  	=& 2T\wh\z\cR^{\text{iso}}(t',t)-\k^{\text{iso}}(t)\CC^{\text{iso}}(t,t')+\int_0^{t}\de s\,\MM^{\text{iso}}_R(t,s)\CC^{\text{iso}}(s,t')
 		 +\int_0^{t'}\de s\,\left[\GG_C(t,s)+\MM_C^{\text{iso}}(t,s)\right]\cR^{\text{iso}}(t',s)
 \ ,\\
 \wh \z\frac{\partial }{\partial t}\cR^{\text{iso}}(t,t')
 	=& \frac{\d(t-t')}{2}-\k^{\text{iso}}(t)\cR^{\text{iso}}(t,t')+\int_{t'}^{t}\de s\,\MM_R^{\text{iso}}(t,s)\cR^{\text{iso}}(s,t')
 \ ,\\
 \frac{\wh\z}2 \frac{\partial }{\partial t} \D(t,t')
 	=& \frac{T}2 + \frac{\k^{\text{iso}}(t)}2 [ \D_r(t') - \D_r(t) - \D(t,t')] + \int_0^t \de s \, \MM_R^{\text{iso}}(t,s) \frac12 [\D_r(t) - \D_r(t') - \D(s,t) + \D(s,t')]
 \\
		&+ \int_0^{\max(t,t')} \!\!\!\! \de s \, \left[\GG_C(t,s)+\MM_C^{\text{iso}}(t,s)\right] \left[\cR^{\text{iso}}(t,s) - \cR^{\text{iso}}(t',s)\right] - 2 T \wh\z \cR^{\text{iso}}(t',t) \, ,
 \\
 \frac{\wh\z}2 \frac{\partial }{\partial t} \D_r(t)
 	=& \frac{T}2 - \k^{\text{iso}}(t) \D_r(t) + \int_0^t \de s \, \MM_R^{\text{iso}}(t,s) \frac12 [\D_r(t) + \D_r(s) - \D(s,t) ]
 		+ \int_0^t\de s\,\left[\GG_C(t,s)+\MM_C^{\text{iso}}(t,s)\right] \cR^{\text{iso}}(t,s)
 \ .
 \end{split}
\end{equation}
The next step is then to derive the self-consistent equations for the memory kernels.

%-----------------
\subsection{Decomposition of the fluctuating gap}
\label{sec:scalar-formulation-decomposition-gap}

The main idea is that the individual displacements of particles with respect to their `drifted' initial position ${\RR_i'(t)=(\hat{1}+\hat{\gamma}(t)) \RR_i}$ remain small, \textit{i.e.}~${\uu_{i}(t) = \mathcal{O}(1/d)}$, and non-trivial motion occurs only along the `longitudinal' direction characterised by the scalar projection $y(t)\propto \hat\rr_0'(t)\cdot\vv(t)$, the motion in the `transverse' plane of dimension ${(d-1)}$ being essentially diffusive.

We now investigate the longitudinal motion in Eq.~\eqref{eq-coshearing-effective-stoch-processes-bis-bis-w} in the direction of ${\rr_0'(t)}$, recalling that its unitary counterpart is simply
${\hat{\rr}_0'(t) \approx \argp{\hat{1} + \hat{\gamma}(t)} \hat{\rr}_0}$ according to Eq.~\eqref{eq-r0-r0t-unitary-vectors-bis}.
We define
\beq
\label{def-longitud-projection-y-cs}
 y(t) \equiv \frac{d}{\ell} \hat{\rr}_0'(t) \cdot \vv(t) \sim \mathcal{O}(1)\, , \quad y(0)=0  \, .
\eeq
We thus divide the interaction distance into three additive contributions with distinct physical origins:
\beq
\label{eq-def-fluctuating-gap}
 \valabs{\rr(t)} = \valabs{\rr_0'(t) + \vv(t)}
 \approx r_0'(t) + \hat{\rr}_0'(t) \cdot \vv(t) + \frac{ \vv(t)^2 }{2 r_0'(t)}
\eeq
where we used that ${\valabs{\vv(t)} \ll r_0'(t)}$ in high dimension.
We can evaluate these different contributions with the definitions~\eqref{def-longitud-projection-y-cs} and
\beq
\begin{split}
 \textit{(i)} \quad & r_0 \equiv \ell \argp{ 1 + \frac{h_0}{d} }
 \quad \Rightarrow \quad r_0'(t) \stackrel{\eqref{eq-r0-r0t-unitary-vectors-bis}}{\approx}
 \ell \argp{1 + \frac{h_0}{d} + \frac{\gamma(t)}{d} g_1g_2 +  \frac{\gamma(t)^2}{2d} g_2^2 }
 \, ,
 \\
 \textit{(ii)} \quad & \Delta_r(t) \equiv \frac{d}{\ell^2} \moy{\valabs{\uu(t)}^2}
 \quad \Rightarrow \quad \valabs{\vv(t)^2}\simeq\moy{\vv(t)^2} \stackrel{\eqref{eq-def-coshearing-7}}{=}
 		\moy{\uu_1(t)^2} + \moy{\uu_2(t)^2} - 2 \moy{\uu_1(t) \cdot \uu_2(t)} \approx \frac{2 \ell^2}{d} \Delta_r(t)
 \, .
\end{split}
\eeq
In \textit{(i)} we make an assumption about the scaling of the initial pair distribution (which can be shown for an equilibrium distribution to derive from the thermodynamics of the system~\cite{KMZ16,MK16}).
\textit{(ii)} is the rescaled mean-square-displacement (MSD) function in the co-shearing frame, in other words the non-affine MSD function; there, we followed the same argument as in Ref.~\cite[Sec. 5.1.1.]{AMZ19}. 
The random variable ${\vv(t)^2=\valabs{\uu_1(t)-\uu_2(t)}^2}$ is a sum over $d$ components dominated by the $\m\geqslant3$ ones which are independent identically distributed variables according to Eq.~\eqref{eq-coshearing-effective-stoch-processes-bis-bis}.
Their large number ${(d-2)}$ makes their contribution dominate the sum ${\vv(t)^2}$ which can be thus replaced by its average, because from the central limit theorem fluctuations are subdominant, as is the crossed term ${\uu_1(t)\cdot\uu_2(t)}$ due to independence of most components.
Gathering all this, we can rewrite Eq.~\eqref{eq-def-fluctuating-gap} as
\beq
\label{eq-def-gap-in-coshearing-frame}
  \valabs{\rr(t)} = \valabs{\rr_0'(t) + \vv(t)}
 	\approx \ell (1 + h(t)/d)
		\, , \quad \text{with} \quad
 	h(t) \equiv h_0 + \gamma(t) g_1 g_2 + \frac12 \gamma(t)^2 g_2^2 + y(t) + \Delta_r(t)
 \,.
\eeq

Physically, the system depends on its initial configuration through the sampling of the initial gap ${h_0}$ and of the initial direction in the shear plane ${g_\mu = \sqrt{d}\, \hat{r}_{0,\mu}}$, ${\mu=1,2}$.
One can show, using hyperpolar angles, that upon integration over the unit sphere, the components $g_\m$ are Gaussian distributed with zero mean and unit variance~\cite[Sec. 3.2]{BU18}. 
Besides, if we start from an equilibrium initial condition (of inverse temperature~$\beta_0$), then $h_0$ is sampled according to ${\mathrm{d}h_0\,e^{h_0-\beta_0 \bar v(h_0)}}$ (see Appendix~\ref{sec-appendix-explicit-coeff}).
Under shear, the system will thus keep a memory of its initial configuration in the shear plane.
Setting ${\gamma(t)=0}$ we recover the results of the isotropic case in the companion paper Ref.~\cite{AMZ19}.

%-----------------
\subsection{Effective Langevin process for the `longitudinal' motion}
\label{sec:scalar-formulation-eff-Langevin-process-longitud}

We now need to write down the effective stochastic process for the fluctuating scalar ${y(t)}$, which is the projection of the relative motion~${\vv(t)}$ on the `drifted' initial distance ${\rr_0'(t)=(\hat{1}+\hat{\gamma}(t)){\rr}_0}$.
Note that the evolution equation of the MSD ${\Delta_r(t)}$, also needed in order to close our effective dynamics, is given by Eq.~\eqref{eq:scaldyneq-copy-from-AMZ19}.

We thus have to project the stochastic process for ${\vv(t)}$, Eq.~\eqref{eq-coshearing-effective-stoch-processes-bis-bis-w}, on ${\hat{\rr}_0'(t)}$.
In Appendix~\ref{sec-projection-eff-stoch-process-detailed} we examine each individual contributions of this projection, taking care of the few non-diagonal coefficients of the kernels.
In fact, all the anisotropic terms turn out to be subdominant in the infinite-dimensional limit, because they involve only two components which are negligible in the $d$-dimensional scalar products.
So, with the rescalings given by Eq.~\eqref{eqC3:fricscal2-shear}, we eventually obtain the following one-dimensional effective stochastic process:
\beq
\label{eq-def-longitudinal-motion-stoch-process}
\begin{split}
 & \widehat{\zeta} \dot y(t)
  = - \kappa^{\text{iso}}(t)  y(t)
 	+ \int_0^t \de s \, \mathcal{M}^{\text{iso}}_R(t,s) \, y(s)
 	- \bar{v}'(h(t))
 	+ \Xi(t)
 \, ,
 \\
 & 	h(t)
 = h_0 + \gamma(t) g_1g_2  + \frac12 \gamma(t)^2 g_2^2 + y(t) + \Delta_r(t)
 \, ,
 \\
 & \text{Initial condition:} \quad y(0)=0 \, , \quad \gamma(0)=0 \, , \quad \Delta_r(0)=0
 \, ,
 \\
 & \text{Gaussian noise:} \quad \moy{\Xi(t)}_\Xi=0 \, ,  \quad \moy{\Xi(t)\Xi(s)}_\Xi= 2T \hat{\zeta} \delta(t-s) +\GG_C(t,s)+ \mathcal{M}^{\text{iso}}_C(t,s)
 \, .
\end{split}
\eeq
The `isotropic' scalar kernels are defined in the section below and in Appendix~\ref{sec-appendix-explicit-coeff}. 
This process is essentially the same as in Ref.~\cite{AMZ19} except for the definition of the gap ${h(t)}$ itself, which is affected by the drift in the shear plane. 
We emphasise that this is the effective stochastic process that characterises the fluctuations along the longitudinal direction ${\rr_0'(t)}$ given essentially by the strained initial condition.

From Eq.~\eqref{eq-def-longitudinal-motion-stoch-process} we can equivalently write an equation directly for the gap in the co-shearing frame:
\beq
\label{eq-effective-stoch-process-gap-MSD-shear}
\begin{split}
 & \widehat{\zeta} \dot h(t)
 = - \kappa^{\text{iso}}(t)  \argp{h(t)-h_0}
 	+ \int_0^t \de s \, \mathcal{M}^{\text{iso}}_R(t,s) \, \argp{h(s) - h_0}
 	- \bar{v}'(h(t)) + \mathcal{B}_{\text{MSD}}(t) + \mathcal{B}_{\text{shear}}(t)
 	+ \Xi(t)
 \, ,
 \\
 & \mathcal{B}_{\text{MSD}}(t)
 \equiv \widehat{\zeta} \dot \Delta_r(t) + \kappa^{\text{iso}}(t) \Delta_r(t) - \int_0^t \de s \, \mathcal{M}^{\text{iso}}_R(t,s) \, \Delta_r(s)
 \, ,
 \\
 & \mathcal{B}_{\text{shear}}(t)
  \equiv \widehat{\zeta} \dot{\gamma}(t) \argp{ g_1g_2 + \gamma(t) g_2^2}
 		+ \kappa^{\text{iso}}(t) \gamma(t) \argp{ g_1g_2 + \frac12 \gamma(t) g_2^2}
 		- \int_0^t \de s \, \mathcal{M}^{\text{iso}}_R(t,s) \gamma(s) \argp{g_1g_2 +  \frac12 \gamma(s) g_2^2 }
 \, ,
 \\
 & \text{Initial condition:} \quad h(0)=h_0 \, , \quad \gamma(0)=0 \, , \quad \Delta_r(0)=0
 \, ,
 \\
 &  \text{Gaussian noise:} \quad \moy{\Xi(t)}_\Xi \, ,  \quad \moy{\Xi(t)\Xi(s)}_\Xi= 2T \hat{\zeta} \delta(t-s)+\GG_C(t,s) + \mathcal{M}^{\text{iso}}_C(t,s)
 \, .
\end{split}
\eeq
In absence of shear we would have ${\mathcal{B}_{\text{shear}}(t)=0}$ and recover the isotropic case results of Ref.~\cite{AMZ19}.
Note that without shear and at equilibrium at temperature ${T=T_0}$ we would also have ${\mathcal{B}_{\text{MSD}}(t)=T}$~\cite{MKZ16,KMZ16}.
%
% So in those cases studying the fluctuations of the projection ${y(t)}$ or of the gap ${h(t)}$ are equivalent, but otherwise the former is more transparent.
%
We denote the statistical averages over the processes in Eq.~\eqref{eq-effective-stoch-process-gap-MSD-shear}, for a given initial condition, as ${\moy{\bullet}_{h \vert h_0,g_1,g_2}}$.

%-----------------
\subsection{Self-consistent definitions of the rescaled kernels}
\label{sec:scalar-formulation-rescaled-kernels}

We can now derive the self-consistent equations for the kernels.
Their rescaled versions~\eqref{eqC3:fricscal2-shear} are calculated in Appendix~\ref{sec-appendix-explicit-coeff} from the high-$d$ limit of Eq.~\eqref{eqC3:Mself}:
\beq
\label{eq-def-kernels-high-dim-gap}
\begin{split}
 \k^{\text{iso}}(t)
 	&= \frac{\wh \f}2 \int^{\infty}_{-\infty}\!\!  \de h_0 \int\DD g_1 \DD g_2 \, e^{h_0} g_{\text{in}}(h_0,g_1,g_2)   \la \redv''(h(t)) + \redv'(h(t)) \ra_{h \vert h_0,g_1,g_2}
 \ , \\
 \MM_C^{\text{iso}}(t,t')
 	&=  \frac{\wh\f}2 \int^{\infty}_{-\infty}\!\!  \de h_0 \int\DD g_1 \DD g_2 \, e^{h_0} g_{\text{in}}(h_0,g_1,g_2)
 		\langle \redv'(h(t)) \redv'(h(t')) \rangle_{h \vert h_0,g_1,g_2}
 \ , \\
 \MM_R^{\text{iso}}(t,s)
 	&=  \frac{\wh\f}2 \int^{\infty}_{-\infty}\!\!  \de h_0 \int\DD g_1 \DD g_2 \, e^{h_0} g_{\text{in}}(h_0,g_1,g_2)  
		\left. \frac{\d \langle \redv'(h(t))  \rangle_{h \vert h_0,g_1,g_2,\PP}}{\d \PP(s)}\right\vert_{\PP=0}
 \ ,
\end{split}
\eeq
where ${\widehat{\varphi}= \rho V_d \ell^d/d}$ is a rescaled packing friction
and ${V_d = \pi^{d/2}/\Gamma(d/2+1)}$ the volume of the unit sphere in $d$ dimensions.
We have to integrate over the initial condition, which we write generally as $g_{\text{in}}(\rr_0)=g_{\text{in}}(r_{0,1},r_{0,2},|\rr_0^\perp|)$, $\rr_0^\perp$ being the component transverse to the shear plane, which is assumed isotropic.
In high dimension, this depends only on the rescaled gap $h_0$ and components $g_1$, $g_2$  (see Appendix~\ref{sec-appendix-explicit-coeff-single-double-deriv}):
${\mathcal{D}g_{\mu}}$ is the measure of the projection of the initial condition on the shear plane, with ${g_\mu=\sqrt{d}\,\hat{r}_{0,\mu}}$ and ${\mathcal{D}g_\mu = \de g_\mu \,e^{- g_\mu^2/2}/\sqrt{2\pi}}$.
The perturbation ${\PP(t)}$ acts in ${\redv'(h(t)) \to \redv'(h(t) - \PP(t))}$, similarly to the definition of the matricial response memory kernel~\eqref{eqC3:Mself}.
These kernels are essentially the same as for the isotropic case (see Ref.~\cite[Sec. 5.1.3]{AMZ19}), except for the integration over the initial condition in the shear plane, which is included in the gap~${h(t)}$.
If we start from equilibrium at inverse temperature ${\beta_0}$, we have ${g_{\text{in}}(h_0,g_1,g_2) = g_{\text{eq}}(h_0)=e^{-\beta_0 \bar{v}(h_0)}}$, which depends only on the radial gap by isotropy of the pair potential.

%-----------------
\subsection{Wrap-up}
\label{sec-scalar-formulation-wrapup}

The isotropy is  broken in presence of shear, as we can see in  the vectorial equations.
Consequently, any observable involving a correlation between the directions ${\mu=1,2}$ might have an explicit dependence on the accumulated shear strain~$\gamma(t)$, as for instance the shear stress ${\sigma(t)=-\Pi_{12}(t)}$~\cite{AMZ19}.
However, because the shear impacts only two directions in a $d$-dimensional space, one can still recover a simple scalar stochastic process ${y(t) \propto \rr_0'(t) \cdot \vv(t)}$ controlling the dynamics, once we have placed ourselves in the co-shearing frame and projected the motion onto the initial direction ${\rr_0}$ of the interaction; all the arguments presented in Ref.~\cite{AMZ19} (in absence of shear), based on the high-dimensional physics, remain thus valid. 
We emphasise that this scalar process depends on shear only through the rescaled gap
${h(t) =h_0 + \gamma(t) g_1g_2  + \frac12 \gamma(t)^2 g_2^2 + y(t) + \Delta_r(t)}$,
keeping a memory of the initial condition via $\arga{h_0,g_1,g_2}$. The scalar effective process is given in Eq.~\eqref{eq-effective-stoch-process-gap-MSD-shear}.
From the scalar effective process one can obtain self-consistently the memory kernels according to Eq.~\eqref{eq-def-kernels-high-dim-gap} and the correlation functions using Eq.~\eqref{eq:scaldyneq-copy-from-AMZ19}.

Once the memory kernels are obtained, one can derive the time evolution of any observable from the effective process, as explained in Ref.~\cite[Sec.7]{AMZ19}.
Examples of one-time quantities are the energy, pressure, and shear stress,
\beq\label{eq:obser}
\begin{split}
 \wh e(t)
 	&=\frac{e(t)}d
 	= \frac{ \wh \f}2 \int^{\infty}_{-\infty}\!\!  \de h_0 \int\DD g_1 \DD g_2 \, e^{h_0} g_{\text{in}}(h_0,g_1,g_2)     \la \redv(h(t)) \ra_{h \vert h_0,g_1,g_2}
 \ , \\
\wh p(t) &= \frac{ \b P(t) }{d \, \r} =
\frac{\b\Tr \hat \Pi(t)}{d^2\, \r}
 	=  - \frac{ \wh \f}2 \int^{\infty}_{-\infty}\!\!  \de h_0 \int\DD g_1 \DD g_2 \, e^{h_0} g_{\text{in}}(h_0,g_1,g_2)  
  \la \b\redv'(h(t)) \ra_{h \vert h_0,g_1,g_2}
 \ , \\
\wh \s(t) &= \frac{ \b \s(t) }{d \, \r} =
-\frac{\b \Pi_{12}(t)}{d\, \r}
 	=   \frac{ \wh \f}2 \int^{\infty}_{-\infty}\!\!  \de h_0 \int\DD g_1 \DD g_2 \, e^{h_0} g_{\text{in}}(h_0,g_1,g_2)  g_1 g_2
  \la \b\redv'(h(t)) \ra_{h \vert h_0,g_1,g_2}
 \ ,
\end{split}
\eeq
and correlation functions can be computed along the same lines~\cite{MKZ16,KMZ16}.

Note that one may restore all the single-particle terms from the initial Langevin dynamics~\eqref{eqC3:GENLang}, namely a finite mass $m$ and non-local friction kernel ${\Gamma_R(t,s)}$.
We refer the reader to the companion paper~\cite{AMZ19}, and more specifically to its summary section~7, because the corresponding equations can be deduced straightforwardly from there.
Note that in presence of a finite mass, the initial distribution of velocities should also be specified, and a kinetic term should be added to some observables.

%_____________________________________________________________
\section{State-following protocol under a finite shear strain}
\label{sec-state-following}

As an application of the effective dynamics in infinite dimension, we want to follow an initial dynamically arrested (\textit{i.e.}~solid) equilibrium state under a quasistatic strain.
The general setting for this protocol has been discussed in Refs.~\cite{CK00,Cu02}, and in the companion paper Ref.~\cite[Sec.~8.2]{AMZ19}, and we briefly recall it below.
We consider that the system is initially prepared in equilibrium below the dynamical transition temperature, where diffusion is arrested and the system is trapped into an infinitely long-lived glassy metastable state.
Correspondingly, the equilibrium dynamics displays a fast time scale,
during which the system ergodically explores the metastable state in equilibrium with its thermal bath, and a formally infinite time scale, during which the system is confined into the metastable state.
As a consequence, the memory kernels do not fully decorrelate and reach a finite value (or `plateau') at long times.
Because the system is in equilibrium at short times, in the long-time limit a slowly applied strain (on a finite timescale ${\tau}$) should be equivalent to an instantaneous 
one ${\dot{\gamma}(t)=\gamma \delta(t)}$ (formally corresponding to $\tau\to 0$).
The latter case has been studied using a static formalism~\cite{YM10,Yo12,RUYZ15}, whereas the former case provides a good starting point for a dynamical justification of the previous static results.

%-----------------
\subsection{Case of a slowly applied strain}
\label{sec-state-following-slowly-applied-strain}

We first consider the case of a strain $\gamma$ smoothly applied over a finite timescale $\tau$. %, and remaining constant after that.
In other words, we assume that ${\hat{\dot{\g}}(t) \to 0}$ for ${t\to\io}$, and therefore ${\hat\g(t) \to \hat\g}$
and we conveniently introduce the following notation:
\beq
 \hat{S}_{\gamma} = \hat{1}+ \lim_{t\to\infty} \hat{\gamma}(t)
 \quad \Rightarrow \quad
 \lim_{t\to\infty} \rr_0'(t)=  \hat{S}_{\gamma}  \rr_0 = \rr_0 + \gamma r_{0,2} \hat{\xx}_1
 \, .
\eeq
If the system is initially prepared in equilibrium in a dynamically arrested phase for ${T<T_d}$, where $T_d$ is the dynamical glass transition temperature,
it will stay in a solid phase for small enough $\g$~\cite{RUYZ15}.
We can thus consider the same restricted equilibrium ansatz as in Ref.~\cite[Sec.~8.2.1]{AMZ19}.
The ansatz will cease to be valid when the system actually `breaks' at the so-called `yielding transition'~\cite{RUYZ15}, beyond which we expect flowing (\textit{i.e.}~diffusive) behaviour.
As we know that in the limit ${d\to\io}$ the anisotropic contributions for the interparticle distance $\vv(t)$ are washed out and survive only within the pair potential (see Eq.~\eqref{eq-def-longitudinal-motion-stoch-process}), for convenience we shall use the vectorial formulation discarding from the start these anisotropic contributions.

At long times ${t,s \to \infty}$ we assume that the memory kernels and the noise can be decomposed as follows~\cite{CK00,Cu02,MKZ16,AMZ19}:
\beq
\label{eq-restricted-equilib-ansatz}
\begin{split}
 & M_C^{\text{iso}}(t,s)
 	\stackrel{(t,s\to\io)}{\longrightarrow} M_f(t-s) + M_\io 
 \ , \\
 & M_R^{\text{iso}}(t,s)
	 \stackrel{(t,s\to\io)}{\longrightarrow} \b \th(t-s)\partial_s M_f(t-s)
 \ , \\
 & \bm{\X}(t)
 	\stackrel{(t\to\io)}{\longrightarrow}  \bm{\X}^f(t) + \bm{\X}^\io 
  \quad \text{with} \quad \left\lbrace \begin{array}{l}
 	\moy{ \X^f_{\m}(t)}_{\text{req}, \bm{\X}^\infty} = 0
 	\, , \quad
 	\moy{ \X^f_{\m}(t) \X^f_{\n}(s) }_{\text{req}, \bm{\X}^\infty}
 		= \d_{\m\n} \left[ T \z \d(t-s) + \frac12 M_f(t-s) \right]
 	\ , \\ \\
 	\overline{ \X^\io_{\m}}=0
 	\, , \quad
 	\overline{ \X^\io_{\m} \X^\io_{\n}}
	 	= \d_{\m\n}  \frac12 M_\io 
 	\ ,
 \end{array} \right.
 \\
 & k^{\text{iso}}(t)
	 \stackrel{(t\to\io)}{\longrightarrow} k_{\infty}
	 \, , \quad
	 k_{\rm eff} \equiv k_\io - \b M_f(0)
 \ .
\end{split}
\eeq
This ansatz assumes a clear timescale separation, with a part related to fast equilibrium fluctuations characterised by the
 fluctuation-dissipation (FDT) relation, and a plateau at long times, encoding the metastability of the glass.
First, $M_\io$ is the long-time plateau of the memory function,
and ${M_f(t-s)}$ is the fast part that decays quickly to zero for ${|t-s|\to\io}$ and is related by FDT to the response ${M_R^{\text{iso}}(t-s)}$ (which has no slow part).
Secondly, the Gaussian noise ${\bm{\X}(t)}$ is decomposed into two Gaussian components, with their corresponding statistical averages: ${\moy{\bullet}_{\text{req}, \bm{\X}^\infty}}$ is an equilibrium average restricted to a given value ${\bm{\X}^\infty}$, and ${\overline{\, \bullet \,}}$ denotes the `disorder average' over ${\bm{\X}^\io}$ (which characterises the glassy metabasin in which the system is trapped).
Thirdly, the kernel $k^{\text{iso}}(t)$ is a single-time quantity and as such it simply goes to its long-time limit value $k^{\text{iso}}_{\infty}$.

Physically, this ansatz expresses the assumption that the system reaches a stationary state at long times (hence, time-translational invariance holds), 
characterised by restricted equilibrium (hence, FDT holds) and by a memory of the initial condition due to the system being trapped in a glass basin (hence, an infinite plateau of the memory function).

With our following assumptions, \emph{the sole impact of a finite shear ${\gamma}$ will be a drift modification of the initial condition.}
Indeed, plugging the ansatz~\eqref{eq-restricted-equilib-ansatz} into the stochastic process for ${\vv(t)}$ given by Eq.~\eqref{eq-coshearing-effective-stoch-processes}, we obtain for long times $t$:
\beq
% \begin{split}
 \frac\z2 \dot \vv(t)
	= 	- \frac{k_\io}2 \vv(t) %- \underbrace{\zeta \hat{\dot{\gamma}}(t)}_{=0} \vv(t)
		+\frac\b2 M_f(0) \vv(t)
		- \frac\b2 \int_0^t \de s \, M_f(t-s) \dot\vv(s)
		-\nabla v (\hat{S}_{\gamma}\rr_0 + \vv(t))
		+ \bm{\X}^f(t) + \bm{\X}^\io \, ,
	\\
% 	  & \meevid{+ (\dots \text{ non-isotropic contributions } \dots)}
% 	\, .
% \end{split}
\eeq
where, as mentioned above, we neglected some anisotropic contributions that vanishes for high $d$. 
Then we can rewrite the stochastic process by regrouping the dissipative terms \textit{versus} the terms deriving from a potential as
\beq
\label{eq-state-following-rewriting-vv}
 \frac\z2 \dot \vv(t) + \frac\b2 \int_0^t \de s \, M_f(t-s) \dot\vv(s)
 = - \argc{ \nabla v (\hat{S}_{\gamma}\rr_0 + \vv(t)) + \frac12 k_{\rm eff} \vv(t) - \bm{\X}^\io} + \bm{\X}^f(t)
 \ .
\eeq
This equation describes an equilibrium dynamics in the modified potential
\beq
\label{eq:veffdef}
 v_{\text{eff}}(\vv)
 	= v(\hat{S}_{\gamma}\rr_0 + \vv) + \frac{k_{\rm eff}}{4} \argp{\vv - 2 \frac{\bm{\X}^\io }{ k_{\rm eff}}}^2  \ ,
\eeq
and has thus the following (normalised) probability distribution~\cite{AMZ19}:
\beq
\label{eq:Peqr}
 P_{\rm req}(\vv;\rr_0,k_{\rm eff}, \bm{\X}^\io)
 	=  \argp{ \frac{\b k_{\rm eff}}{4\pi} }^{d/2} \frac{ e^{-\b v(\hat{S}_{\gamma}\rr_0 + \vv) - \frac{\b k_{\rm eff}}4 (\vv - 2 \bm{\X}^\io/k_{\rm eff})^2 } }
{ e^{\frac{1}{\b k_{\rm eff}}\Lap   } e^{-\b v(\hat{S}_{\gamma}\rr_0 + 2  \bm{\X}^\io/k_{\rm eff})}}
\ ,
\eeq
\emph{provided that ${k_{\rm{eff}} >0}$} for the potential to be confining, allowing for a restricted equilibrium distribution.
Note that we used a compact notation for a Gaussian convolution 
% \footnote{\beq e^{\frac{A}{2} \nabla^2} f(\rr)=\int \de \xx\,\frac{e^{-\frac{\xx^2}{2A}}}{(2\p A)^{d/2}} f(\xx+\rr)
% \eeq} 
via the operator ${e^{\frac{A}{2} \nabla^2}}$~\cite[Sec.~8.2.1]{AMZ19}.

Consistently, we can then assume that the long-time MSDs have a finite limit, in other words that the particles are typically stuck into a metabasin (the above confining potential).
The relation between the long-time MSDs and the memory kernels are the same as those given in Ref.~\cite{AMZ19}:
\beq
\label{eq:MSDslongtime}
 \DE_r
 	= \frac1d \overline{  \la | \uu |^2 \ra_{\rm req}}
 	= \frac{1}{\b k_{\rm eff}} + \frac{M_\io}{k_{\rm eff}^2} 
 \ , \quad
 \DE
 	= \frac2d \overline{ [  \la | \uu |^2 \ra_{\rm req} - | \la \uu \ra_{\rm req} |^2 ] }
	=  \frac{2}{\b k_{\rm eff}}
 \ .
\eeq
Note that these MSDs are defined within the co-shearing frame (\textit{i.e.}~removing the affine contribution to the displacement), but we skipped the exponent `cs'.
Introducing for convenience
${\AE = \overline{| \la \uu \ra_{\rm req} |^2} = 2 \DE_r - \DE = 2 M_\io/k_{\rm eff}^2}$, 
one can therefore express the self-consistent equations as follows:
\begin{eqnarray}
\label{eq-state-following-equa1}
 \frac1{\DE} - \frac{\AE}{\DE^2}
 	&=& \frac{\b}2 ( k_{\rm eff} - \b M_\io )
 	= -\frac{ \r}{d} \int \de\rr_0 \, e^{-\Bi v(\rr_0)}  e^{\frac{\AE}{2}\Lap }
	\left[ \frac{ \frac{\Lap}2 e^{\frac{\DE}{2}\Lap   }  e^{-\b v(\hat{S}_{\g}\rr_0 )} }
{ e^{\frac{\DE}{2}\Lap   } e^{-\b v(\hat{S}_{\g}\rr_0 )}}  \right]
\ ,
\\
\label{eq-state-following-equa2}
 \frac1{\DE}
	&=&	\frac{\b k_{\rm eff}}2
	= -\frac{\r}{d} \int \de\rr_0 \, e^{-\Bi v(\rr_0)} e^{\frac{\AE}{2}\Lap   } \frac{\Lap}2\log
			\argc{ e^{\frac{\DE}2 \Lap} e^{-\b v(\hat{S}_{\g}\rr_0)} }
 \, .
\end{eqnarray}
From these two last equations one can access ${\lbrace \DE,\AE \rbrace}$ and thus the long-time limit MSDs ${\lbrace \DE, \DE_r \rbrace}$ or,
 equivalently, ${\lbrace k_{\rm eff},M_\infty \rbrace}$.
Note that Eqs.~\eqref{eq-state-following-equa1}-\eqref{eq-state-following-equa2} 
are obtained by taking the derivatives with respect to $\AE$ and $\DE$ of the glass free energy 
\beq
\label{eq-state-following-free-energy-strain}
 -\b {\rm f}_g
 	= \frac{d}2 \left[ 1 + \log(\pi \DE) + \frac{\AE}{\DE} \right] + \frac\r2 \int \de \rr_0 \, e^{-\Bi v(\rr_0)} e^{\frac{\AE}2 \Lap} \log \left[ e^{\frac{\DE}2 \Lap} e^{-\b v(\hat S_{\g} \rr_0)} \right]
\ .
\eeq
The control parameter of Eqs.~\eqref{eq-state-following-equa1}-\eqref{eq-state-following-equa2} are the two inverse temperatures ${\beta}$ and $\beta_0$, the specific interaction potential ${v(\valabs{\rr})}$ and the density~$\rho$.
The finite strain is present only through the factors ${e^{-\beta v(\hat{S}_\gamma \rr_0)}}$.
Physically, in the range of parameters for which these two equations can be satisfied, the system can sustain a solid phase after a finite shear strain, characterised by finite values for ${\lbrace k_{\rm eff}, M_\infty , D_r,D\rbrace}$.
On the contrary, when these equations cannot be satisfied, that means that our previous assumptions break down, and we should expect to have ${\lbrace k_{\rm eff}=0, M_\infty=0 \rbrace}$ and diverging long-time MSDs.
Physically, this corresponds to the flow regime beyond the yielding transition.

%-----------------
\subsection{Case of an instantaneous strain}
\label{sec-state-following-instantaneous-strain}

We now consider the application of a static shear strain, which formally corresponds to ${\dot\g(t) = \g \d(t)}$.
In other words, we consider an equilibrium initial configuration $\{ \RR_i \}$,
to which we immediately apply an affine strain
${\RR_i \to \RR_i' = \RR_i + \g R_{i2} \hat\xx_1 = \hat S_\g \RR_i}$,
where ${\hat S_\g = \hat1 +  \g \hat\xx_1 \hat\xx_2^T}$,
and we then run the dynamics starting from the sheared initial condition,
as it is done in the static approach of Refs.~\cite{YM10,Yo12,RUYZ15}.

This amounts to first sampling the initial condition according to the radial 
distribution ${e^{-\beta_0 v(\rr_0)}}$, and then
using the affinely deformed ${\rr_0' = \hat S_\g \rr_0}$ as the initial condition 
of the effective process for ${\vv(t)}$.
Because in this case the deformation only appears in the initial condition, we can use the dynamical equations without strain discussed in Ref.~\cite{AMZ19},
simply using as initial condition a random vector $\rr_0'$ such that $\rr_0 = \hat S_{-\g} \rr_0'$ is sampled in equilibrium, \textit{i.e.}~$\rr_0'$ is sampled with distribution $e^{-\beta_0 v(\hat S_{-\g} \rr_0)}$.
The glass free energy then corresponds to the one in Ref.~\cite[Eq.(140)]{AMZ19},
with a modified distribution of the initial condition, ${e^{-\beta_0 v( \rr_0)}\to e^{-\beta_0 v(\hat S_{-\g} \rr_0)}}$, leading to
\beq
\label{eq-state-following-free-energy-strain2}
 -\b {\rm f}_g
 	= \frac{d}2 \left[ 1 + \log(\pi \DE) + \frac{\AE}{\DE} \right] + \frac\r2 \int \de \rr_0' \, e^{-\Bi v(\hat S_{-\g} \rr_0')} e^{\frac{\AE}2 \Lap} \log \left[ e^{\frac{\DE}2 \Lap} e^{-\b v(\rr_0')} \right]
\ .
\eeq
Changing variables back to ${\rr_0 = \hat S_{-\g} \rr_0'}$, and 
exploiting the fact that
\textit{(i)}~the Jacobian is unity since ${\det \wh S_\g=1}$, and
\textit{(ii)}~for ${d\to\io}$ one can consider $\wh S_\g$ as unitary to leading order, thus ${\nabla_{\rr_0'}^2\simeq \nabla_{\rr_0}^2}$,
one can show that Eq.~\eqref{eq-state-following-free-energy-strain2} is equivalent to Eq.~\eqref{eq-state-following-free-energy-strain}.

Both Eqs.~\eqref{eq-state-following-free-energy-strain} and~\eqref{eq-state-following-free-energy-strain2} reproduce, in the high-dimensional limit, the result derived in Ref.~\cite{RUYZ15}.
We only give here the main idea of the proof~\cite{BU18}.
On the one hand, in the ideal gas term of the free energy one should simply rescale the MSDs appropriately, ${\DE = \D \ell^2/d^2}$ and ${\AE = \a \ell^2/d^2}$.
On the other hand, for the excess free energy term given in Eq.~\eqref{eq-state-following-free-energy-strain2}, in high dimension the convolution becomes
${ e^{\frac{\DE}2 \Lap} e^{-\b v(\rr_0')}
 	\to  e^{\frac{\D}2 \frac{\de^2}{\de h^2}}e^{-\b\redv(h_0+\D/2)}
}$~\cite{PZ10}.
Next, similarly to the discussion in Sec.~\ref{sec:scalar-formulation-decomposition-gap} leading to Eq.~\eqref{eq-def-gap-in-coshearing-frame},
one observes that in high dimension
${|\hat S_{-\g} \rr_0'| =  \ell \left(1 + h_0/d - \g \hat r_{0,1} \hat r_{0,2} + \frac{\g^2}2 \hat r_{0,2}^2\right)}$ 
and the components of ${\sqrt{d}\, \hat \rr_0}$ become uncorrelated random Gaussian variables $g_\m$ with zero mean and unit variance.
So, integrating by parts the operator $e^{\frac{\AE}2 \Lap}$ and replacing vectorial integration by scalar ones in high $d$, we finally obtain
\beq
\begin{split}
 -\b {\rm f}_g^{\rm ex}
 	& =  \frac\r2 \int \de \rr_0' \, \left[ e^{\frac{\AE}2 \Lap}e^{-\Bi v(\hat S_{-\g} \rr_0')}\right]  \log \left[ e^{\frac{\DE}2 \Lap} e^{-\b v(\rr_0')} \right]
 \\
 	&= \frac{d \wh \f}2 \int_{-\io}^\io \DD g_1 \DD g_2 \, \de h_0 \, e^{h_0} \, 
	\left[  e^{\frac{\a}2 \frac{\de^2}{\de h^2}}e^{-\b\redv(h_0+\a/2 - \g g_1 g_2 + \g^2 g_2^2/2)} \right]  \log \left[  e^{\frac{\D}2 \frac{\de^2}{\de h^2}}e^{-\b\redv(h_0+\D/2)} \right]
 \\
	&= \frac{d \wh \f}2 \int_{-\io}^\io \de h_0 \, e^{h_0} \, \left[ \int \DD g  \,
	 e^{\frac{\a + \g^2 g^2}2 \frac{\de^2}{\de h^2}}e^{-\b\redv\left(h_0+\frac{\a+\g^2 g^2}2 \right)} \right]  \log \left[  e^{\frac{\D}2 \frac{\de^2}{\de h^2}}e^{-\b\redv(h_0+\D/2)} \right]
\ .
\end{split}
\eeq
This results coincides with the replica-symmetric result of Ref.~\cite{RUYZ15}.

Similarly to the case of random forces, it has been shown in Ref.~\cite{RUYZ15} that applying a small strain $\g$ to a solid solution (with finite $\DE$), the solution is first weakly modified, thus describing the elasticity of the solid, while for larger $\g$ it disappears at the yielding point, beyond which diffusive dynamics should set in.

%_____________________________________________________________
\section{Conclusions}
\label{sec-conclusions}

In this paper, we derived the dynamical mean field equations that describe the dynamics of an infinite-dimensional particle system submitted to a shear strain.
They are formulated in terms of a scalar stochastic process, Eq.~\eqref{eq-effective-stoch-process-gap-MSD-shear}, whose memory kernels are determined self-consistently by Eq.~\eqref{eq-def-kernels-high-dim-gap}, with correlation functions given by Eq.~\eqref{eq:scaldyneq-copy-from-AMZ19}.
Physical observables are obtained as averages over the effective process, such as Eq.~\eqref{eq:obser}.
For the derivation, we used the same key features of high-dimensional physics as in Refs.~\cite{MKZ16,Sz17,AMZ19}, but only once we moved into the co-shearing frame:
\textit{(i)}~small displacements ${\uu_i(t)}$ around the drifting initial positions ${\RR_i'(t)=\RR_i + \gamma(t) R_{i,2} \hat\xx_1}$,
\textit{(ii)}~uncorrelated numerous neighbours,
and \textit{(iii)}~a statistical isotropy of the system in the ${(d-2)}$ directions orthogonal to the shear plane.

The derivation presented in this paper is based on a perturbative cavity approach~\cite{MPV87}.
As in the companion paper~\cite{AMZ19}, it is possible to derive the same set of dynamical equations through a MSRDDJ path-integral formalism.
The essential logical steps,  as in Secs.~4 and 5.1 of the companion paper~\cite{AMZ19}, are the following:
\begin{itemize}

\item[\textit{(i)}] one must write the dynamical action in the co-shearing frame as in Eq.~\eqref{eqC3:GENLang-for-ucs};
 the dynamical virial expansion can then be truncated to its first excess term;

\item[\textit{(ii)}] the excess term depends only on a stochastic variable ${\vv(a)=\uu_1(a)-\uu_2(a)}$ (in the supersymmetric formulation), 
which, as other dynamical quantities, can be treated as isotropic out of the shear plane.
Following similar arguments than the ones presented in Ref.~\cite[Sec.~5.1.1]{AMZ19} and Sec.~\ref{sec:scalar-formulation-decomposition-gap}, one obtains a scalar process for the 
projection ${y(a)=\frac{d}{\ell}\hat\rr_0'(t)\cdot\vv(a)}$, where the only modifications with respect to Ref.~\cite{AMZ19} are relative to the measure of the initial condition and the dependence upon $\{h_0,g_1,g_2\}$ inside the rescaled pair potential as in Eq.~\eqref{eq-def-gap-in-coshearing-frame};

\item[\textit{(iii)}] the ideal gas term can be treated in an identical way to Ref.~\cite{AMZ19}, taking into account that the dominant contribution 
of tensors come from the diagonal;

\item[\textit{(iv)}] finally the saddle-point equation is in principle matricial but its diagonal defines the `isotropic' kernels of Eq.~\eqref{eq-def-kernels-high-dim-gap}, retrieving the correct dynamics for 
the projected motion as in Eq.~\eqref{eq-def-longitudinal-motion-stoch-process}.
\end{itemize}

We believe that the programme, initiated in Ref.~\cite{KMZ16}, of establishing dynamical mean field equation to describe, in full generality, the out-of-equilibrium dynamics of infinite-dimensional particle systems, is now complete.
As a side result, this study provides a fully dynamical derivation of the state following equations obtained in Ref.~\cite{RUYZ15} via the replica method.
These equations were used in Refs.~\cite{RUYZ15,RU16,UZ17} to provide physical predictions for the quasistatic rheology of strained glasses, in good qualitative agreement with numerical simulations in $d=3$~\cite{NYZ16,JY17,JUZY18}.

The challenge for the future is of course to extract physical information from these equations.
While fully analytical solutions seem to be out of reach, it might be possible to develop approximations schemes to map the complete equations on simplified models,
such as elasto-plastic models~\cite{NFMB19}. This would provide an interesting first-principle derivation of such models, and clarify the physical assumptions involved
in their formulation. Complementarily,
methods to solve these equations numerically are currently being developed~\cite{RBBC19}.
This requires, starting from an ansatz for the memory kernels, to perform the following steps: {\it (i)} generate several realisations of correlated noise, {\it (ii)} solve numerically a one-dimensional stochastic equation with such colored noise, 
and {\it (iii)} determine a new estimate for the memory kernels by measuring the appropriate correlation functions on the stochastic trajectories. The procedure is then repeated until convergence~\cite{RBBC19}.
In principle, straightforward application of these numerical methods might allow one to extract the physical behaviour in several interesting regimes, such as equilibrium dynamics, start-up shear strain dynamics, dynamics of active matter, micro-rheology, critical dynamics close to jamming, and so on.
In situations where a strong separation of time scales is observed, such as aging, the numerical solution will be more challenging and analytical insight will be necessary~\cite{CK93,FFR19}.
The study of out-of-equilibrium stationary states could also be possible by self-consistently determining the distribution of initial conditions, ${g_{\rm in}(h_0,g_1,g_2)}$.

Another important direction for future research is to go beyond the pure mean field description by including corrections of different nature, such as the static structure of the fluid and
the wave-vector dependence of dynamical correlators.
These features can be included by resummations of the high $d$ expansion~\cite{hansen}.
They are not expected to affect qualitatively the dynamics, but will increase the quantitative accuracy of the theory, as in MCT~\cite{Go09}.
Other important finite-dimensional ingredients are, for example, the spatial fluctuations of dynamical observables such as the shear stress, that are often treated phenomenologically, \textit{e.g.}~in elasto-plastic models~\cite{NFMB19}.
It would be nice to construct a theory partially based on the dynamical mean field equations, to capture at least some of the ingredients that are at the basis of such models.

%_____________________________________________________________
\section*{Acknowledgments}

This paper and its companion~\cite{AMZ19}, are dedicated to Giorgio Parisi in his 70th birthday; Giorgio is a great source of inspiration for all of us and discussions with him, as well as previous joint work, have been crucial for the development of this research.
We would also like to thank Giulio Biroli, Matthias Fuchs, Jorge Kurchan, Alessandro Manacorda, Grzegorz Szamel and Pierfrancesco Urbani
for fruitful discussions related to this work.
This project has received funding from the European Research Council (ERC) under the European Union Horizon 2020 research and innovation programme (grant agreement n. 723955 - GlassUniversality).
This research was supported in part by the National Science Foundation under Grant No. NSF PHY-1748958.
E.A. acknowledges support from the SNSF Ambizione Grant PZ00P2{\_}173962.
T.M. acknowledges funding from the Grant ANR-16-CE30-0023-01 (THERMOLOC).

%_____________________________________________________________

\appendix

%_____________________________________________________________
\section{Explicit anisotropic coefficients of the memory kernels}
\label{sec-appendix-explicit-coeff}

We provide thereafter the explicit expressions for the non-zero coefficients of the kernels and overall drift,
self-consistently defined in Eq.~\eqref{eqC3:Mself} by the high-dimensional effective stochastic process~${\vv(t)}$.

%-------
\subsection{Derivatives and integrations in infinite dimension}
\label{sec-appendix-explicit-coeff-single-double-deriv}

Because all the coefficients are statistical averages of simple or double derivatives of the interaction potential, we will use Eq.~\eqref{eq-derivatives-radial-functions-refs}, which can be simplified for high $d$ through Eq.~\eqref{eq-r0-r0t-unitary-vectors-bis} and the scaling of the potential in Eq.~\eqref{eq-def-rescaled-potential-1}:
\beq
\label{eq-double-derivative-force-bis-bis}
\begin{split}
 \nabla_1^2 v \argp{\valabs{{\rr}_0'(t) + {\vv}(t)}}
  & \simeq
  	\frac{d}{\ell^2}\argc{ \bar{v}''(h(t)) \argp{g_1+\g(t)g_2}^2 + \bar{v}'(h(t))}
 \, ,
 \\
%  \nabla_1^2 v \argp{\valabs{{\rr}_0'(t) + {\vv}(t)}}
%   & \approx
%   \frac{d}{\ell^2}\argc{ \bar{v}''(h(t)) \argp{g_1^2+\gamma(t)^2g_2^2} + \bar{v}'(h(t))} + \frac{d}{\ell^2} \bar{v}''(h(t)) \, 2 \gamma(t) g_1g_2
%  \, ,
%  \\
 \nabla_1 \nabla_2 v \argp{\valabs{{\rr}_0'(t) + {\vv}(t)}}
  & =\nabla_2 \nabla_1 v \argp{\valabs{{\rr}_0'(t) + {\vv}(t)}}
  \simeq\frac{d}{\ell^2} \bar{v}''(h(t)) \argc{g_1g_2+\gamma(t) g_2^2}  \, 
 \, ,
   \\
 \nabla_\mu \nabla_\nu v \argp{\valabs{{\rr}_0'(t) + {\vv}(t)}}
 	& \stackrel{(\mu,\nu \neq 1)}{\simeq}\frac{d}{\ell^2}\argc{ \bar{v}''(h(t))g_{\mu} g_{\nu}
  	+ v'(h(t)) \delta_{\mu\nu}}
 \, .
\end{split}
\eeq
Concerning the integration measure, one has ${\de\rr_0=r_0^{d-1}\de r_0\,\de \hat\rr_0}$ with ${r_0=\ell(1+h_0/d)}$.
In high $d$, the radial part becomes ${r_0^{d-1}\de r_0 = \ell^d (1+h_0/d)^{d-1} \de h_0/d \sim (\ell^d/d) e^{h_0} \de h_0}$.
Furthermore, as shown in Ref.~\cite[Sec. 3.2]{BU18}, in high $d$ the solid angle measure reads ${\de\hat\rr_0\approx \Omega_d \prod_{\m=1}^d\DD g_\m}$ with ${\Omega_d = d V_d}$ the $d$-dimensional solid angle, ${g_\m = \sqrt{d} \hat r_{0,\m}}$ and ${\DD g}$ the Gaussian measure with unit variance and zero mean.
Hence, in high $d$,
\beq
\de\rr_0\approx V_d \ell^d e^{h_0} \de h_0 \prod_{\m=1}^d\DD g_\m \ ,
\qquad
g_\m = \sqrt{d} \, \hat r_{0,\m} \approx \sqrt{d} \, r_{0,\m}/\ell \ , \qquad r_0=\ell(1+h_0/d) \ .
\eeq
The initial condition, here and in the following, is assumed to be isotropic out of the shear plane, \textit{i.e.}~${g_{\text{in}}(\rr_0)=g_{\text{in}}(r_{0,1},r_{0,2},|\rr_0^\perp|)}$ where the transverse radial component ${|\rr_0^\perp|\sim r_0}$ for high $d$, and ${r_{0,\m}=r_0\hat r_{0,\m}\sim r_0 g_\m/\sqrt d}$.
Therefore, provided the  dependence in these variables is scaled suitably so that it remains finite for high $d$, we get ${g_{\text{in}}(\rr_0)\sim g_{\text{in}}(h_0,g_1,g_2)}$ with a slight abuse of notation.
All in all, we can replace
\beq
\frac{\rho}{d} \int \de \rr_0 \, g_{\text{in}}(\rr_0)\,\bullet
        \ \to\  \wh{\varphi} \int_{-\infty}^{\infty} \de h_0\, e^{h_0}  \int  \prod_{\mu=1}^d \mathcal{D} g_\mu \, g_{\text{in}}(h_0,g_1,g_2)\,\bullet
 \, , \quad
 \moy{\bullet}_{\vv} \to \moy{\bullet}_{h\vert h_0,g_1,g_2}
 \, ,
\eeq
with ${\wh{\varphi}=\rho V_d \ell^d / d}$ the rescaled packing fraction defined in Sec.~\ref{sec:scalar-formulation-rescaled-kernels}.

%-------
\subsection{Overall drift ${{\bf \Xi}(t)}$: average force}
\label{sec-appendix-explicit-coeff-overall-drift-Xi}

The overall drift ${{\bm{\bar\Xi}}(t)}$ has non-zero components only in the shear plane ${\{\hat\xx_1,\hat\xx_2\}}$ by symmetry.
For ${\m=1,2}$ we have
\beq
\begin{split}
 \sqrt2\,\bar\Xi_\mu (t)
	\stackrel{\eqref{eqC3:Mself}}{=}
	& - \r  \int \de \rr_0 \, g_{\text{in}}(\rr_0) \, \moy{ v'\argp{\valabs{\rr_0'(t) + \vv(t)}}}_{\vv} \argp{\hat{r}_{0,\mu} + \gamma(t) \hat{r}_{0,2} \, \delta_{1,\mu}}
 \\
 	\stackrel{(d \to \infty)}{=}
 	& - \frac{d^{\frac32}}{\ell}\wh\f \int_{-\infty}^{\infty} \de h_0\, e^{h_0} \int  \DD g_1\DD g_2 
 \,  g_{\text{in}}(h_0,g_1,g_2)\moy{\bar{v}' (h(t))}_{h \vert h_0,g_1,g_2} \argp{g_\mu + \gamma(t) g_2 \delta_{1,\mu}}
	=\OO(d^{\frac32})\, .
\end{split}
\eeq
which is negligible when compared to $\zeta$, $\hat M_C$ or $\G_C$ as they scale as ${\OO(d^2)}$. As a consequence it can be overlooked in the high-dimensional limit.

%-------
\subsection{Memory kernel ${\hat{M}_C(t,s)}$: force-force correlator}
\label{sec-appendix-explicit-coeff-MC}

The matricial memory kernel ${\hat{M}_C(t,s)}$ is given, before dimensional rescaling, by
\beq
\begin{split}
 & M^{\m\n}_C(t,s)
 \stackrel{\eqref{eqC3:Mself}}{=}
	\r \int \de \rr_0 \, g_{\text{in}}(\rr_0) \, \moy{
		v'\argp{\valabs{\rr_0'(t) + \vv(t)}}v'\argp{\valabs{\rr_0'(s) + \vv(s)}}}_{\vv}
		\argp{\hat{r}_{0,\mu} + \gamma(t) \hat{r}_{0,2} \, \delta_{1,\mu}}
		\argp{\hat{r}_{0,\nu} + \gamma(s) \hat{r}_{0,2} \, \delta_{1,\nu}}
 \, .
\end{split}
\eeq
Because the shear impacts only two directions, we have $M_C^{\m\n}=0$ for ${\m\neq\n\notin\{1,2\}}$ and $M_C^{\m\m}$ is independent of $\m$ for ${\m\neq1,2}$. We can thus perform the following approximation:
\beq
\m\neq1,2\quad\Rightarrow\quad M_C^{\m\m} \stackrel{(d\to\io)}{\simeq}\frac1d\sum_{\n=1}^dM_C^{\n\n}=:M_C^{\text{iso}}\, ,
\eeq
\beq
\begin{split}
 M^{\text{iso}}_C(t,s)
 &	 	\stackrel{(d\to\io)}{=}\r \int \de \rr_0 \, g_{\text{in}}(\rr_0) \, \moy{v' \argp{\valabs{{\rr}_0'(t) + {\vv}(t)}} v' \argp{\valabs{{\rr}_0'(s) + {\vv}(s)}}}_{\vv} \, .
\end{split}
\eeq
The rescaled kernel in the limit ${d\to\io}$ reads:
\beq
\begin{split}
 & \mathcal{M}^{\m\n}_C(t,s)
 \equiv \frac{\ell^2}{2d^2}M^{\m\n}_C(t,s)
 \\
  & \Rightarrow \quad \left\lbrace \begin{array}{rl}
 \mathcal{M}^{11}_C(t,s)
 	&=   \frac{\wh\f}{2} \int_{-\infty}^{\infty} \de h_0\, e^{h_0}  \int  \mathcal{D} g_1 \mathcal{D} g_2
	  \,g_{\text{in}}(h_0,g_1,g_2) \moy{\bar{v}' (h(t)) \bar{v}'(h(s))}_{h \vert h_0,g_1,g_2} \\ 
	  &\qquad	\times  \argc{g_1^2 +(\gamma(t)+\gamma(s)) g_1 g_2  + \gamma(t)\gamma(s) g_2^2} \, ,
 \\
  \mathcal{M}^{12}_C(t,s)
  	&= \frac{\wh\f}{2} \int_{-\infty}^{\infty} \de h_0\, e^{h_0} \int  \mathcal{D} g_1 \mathcal{D} g_2
	  \, g_{\text{in}}(h_0,g_1,g_2)\moy{\bar{v}' (h(t)) \bar{v}'(h(s))}_{h \vert h_0,g_1,g_2} \argp{g_1 g_2 + \gamma(t) g_2^2} \, ,
 \\
  \mathcal{M}^{21}_C(t,s)
  	&= \frac{\wh\f}{2} \int_{-\infty}^{\infty} \de h_0\, e^{h_0}  \int  \mathcal{D} g_1 \mathcal{D} g_2
	  \,g_{\text{in}}(h_0,g_1,g_2) \moy{\bar{v}' (h(t)) \bar{v}'(h(s))}_{h \vert h_0,g_1,g_2} \argp{g_1 g_2 + \gamma(s) g_2^2} \, ,
	  \\
\mathcal{M}^{22}_C(t,s)
   &= \frac{\wh\f}{2} \int_{-\infty}^{\infty} \de h_0\, e^{h_0}  \int  \mathcal{D} g_1 \mathcal{D} g_2
	  \,g_{\text{in}}(h_0,g_1,g_2) \moy{\bar{v}' (h(t)) \bar{v}'(h(s))}_{h \vert h_0,g_1,g_2}g_2^2
 \\ 
  \mathcal{M}^{\mu\mu}_C(t,s)
   &= \mathcal{M}^{\text{iso}}_C(t,s)
   = \frac{\wh\f}{2} \int_{-\infty}^{\infty} \de h_0\, e^{h_0}  \int  \mathcal{D} g_1 \mathcal{D} g_2
	  \, g_{\text{in}}(h_0,g_1,g_2)\moy{\bar{v}' (h(t)) \bar{v}'(h(s))}_{h \vert h_0,g_1,g_2} \quad \text{for } \mu \geqslant3
 \, .
 \end{array} \right.
\end{split}
\eeq
The crucial remark here is that all these kernel components are of the same order when $d\to\io$. But because the one associated to spatial components ${\mu\geqslant 3}$ are much more numerous, the anisotropic components can be neglected everywhere.

%-------
\subsection{Memory kernel ${\hat{M}_R(t,s)}$: average response of the force}
\label{sec-appendix-explicit-coeff-MR}

The matricial response memory kernel ${\hat{M}_R(t,s)}$ is given, before dimensional rescaling, by
\beq
 M^{\m\n}_R(t,s)
 \stackrel{\eqref{eqC3:Mself}}{=}
	\r  \int \de \rr_0 \, g_{\text{in}}(\rr_0) \, \frac{\d \moy{  v'\argp{\valabs{\rr_0'(t) + \vv(t)}} }_{\vv,\bm{P}}}{\d P_{\nu}(s)}\Bigg\vert_{\bm P=\bm0} \argp{\hat{r}_{0,\mu} + \gamma(t) \hat{r}_{0,2} \, \delta_{1,\mu}}
 \, .
\eeq
In the following we shall be less concerned by this actual kernel than by the similar `frictional' term arising from the projection of the force in Eq.~\eqref{eq:forceexpandcavity} (or rather the difference ${\bm F_i(t)-\bm F_j(t)}$ which yields the process for the interparticle distance Eq.~\eqref{eqC3:GENLang-for-wcs-ter}) on the dominant direction of the interparticle force.
This involves computing the perturbation of the force between two particles when the interaction distance is shifted (here the shift at time $s$ is referred to as ${-\bm P(s)}$).
First, let us note that for high $d$ the vector $\moy{  \nabla v\argp{\valabs{\rr_0'(t) + \vv(t)}} }_{\vv,\bm{P}}$ is essentially directed along ${\hat\rr_0'(t)}$.
Second, since the potential is radial, one has to inspect the distance:
\beq
\label{eq:appMR}
\valabs{\rr_0'(t)+\vv(t)-\bm{P}(t)}=r_0'(t)\sqrt{1+2\frac{\hat\rr_0'(t)\cdot\vv(t)}{r_0'(t)}+\frac{\vv(t)^2}{r_0'(t)}-2\bm{P}(t)\cdot\frac{\hat\rr_0'(t)+\frac{\vv(t)}{r_0'(t)}}{r_0'(t)}+\frac{\bm{P}(t)^2}{r_0'(t)^2}} \ .
\eeq
The first $\bm P$-independent terms in the square root have been studied in Sec.~\ref{sec:scalar-formulation-decomposition-gap} and are of~${\OO(1/d)}$.
Similarly, it follows that ${\bm P(t)\propto \hat\rr_0'(t)}$ maximises the impact of the perturbation on the distance and must scale as ${\bm P(t)=\OO(1/d)}$.
Let us then define ${\bm P(t)=(\ell/d)\PP(t)\hat\rr_0'(t)}$, then Eq.~\eqref{eq:appMR} is written as, following the same arguments as in Sec.~\ref{sec:scalar-formulation-decomposition-gap},
\beq
\label{eq:appMR2}
\valabs{\rr_0'(t)+\vv(t)-\bm{P}(t)} \stackrel{(d\to\io)}{=}r_0'(t)+\frac{\ell}{d}\argc{y(t)+\D_r(t)-\PP(t)}  \, ,
\eeq 
up to subdominant corrections.
Incidentally, this shows that such a perturbation corresponds, in the rescaled potential ${\redv(h(t))}$, to the gap shift ${h(t)\to h(t)-\PP(t)}$ defined in Eq.~\eqref{eq-def-gap-in-coshearing-frame}. 
As an example, if one considers instead a perturbation in any fixed direction $\hat\xx_\m$, \ie 
${\bm P(t)=(\ell/d)\PP_\m(t)\hat\xx_\m}$, one would get instead of $\PP(t)$ in Eq.~\eqref{eq:appMR2} a term ${\PP_\m(t)\hat r_{0,\m}}$ which would be subdominant as ${\hat r_{0,\m}=\OO(1/\sqrt d)}$.

From the above discussion we conclude that the matrix ${\d\moy{  \nabla v\argp{\valabs{\rr_0'(t) + \vv(t)}} }_{\vv,\bm{P}}/\d\bm P(s)}$ is dominated by its component along the projector ${\hat\rr_0'(t)\hat\rr_0'(s)^T}$, in the sense that any double contraction of it will be negligible with respect to ${\hat\rr_0'(t)\cdot\d\moy{  \nabla v\argp{\valabs{\rr_0'(t) + \vv(t)}} }_{\vv,\bm{P}}/\d P(s)}$ where ${\bm P(s)=P(s)\hat\rr_0'(s)}$.
This is emphasised by the following definition of a scalar (contracted) kernel (respectively bare and rescaled) corresponding to this projector component:
\beq
\begin{split}
 M^{\text{iso}}_R(t,s)
 &	:=  \frac\r d  \int \de \rr_0 \, g_{\text{in}}(\rr_0) \, \frac{\d \moy{  \hat\rr_0'(t)\cdot\nabla v\argp{\valabs{\rr_0'(t) + \vv(t)}} }_{\vv,\bm{P}}}{\d P(s)}\Bigg\vert_{P=0}
 \quad \text{with }\bm P(s)=P(s)\hat\rr_0'(s)\, ,
 \\
  \mathcal{M}^{\text{iso}}_R(t,s)
 & := \frac{\ell^2}{2 d^2} M^{\text{iso}}_R(t,s)
 	\stackrel{(d \to \infty)}{=} \frac{\wh\f}{2} \int_{-\infty}^{\infty} \de h_0\, e^{h_0}  \int \mathcal{D} g_1 \mathcal{D} g_2
 \,g_{\text{in}}(h_0,g_1,g_2) \frac{\d \moy{  \bar{v}'(h(t))}_{h\vert h_0,g_1,g_2, \PP}}{\d \PP(s)}\Bigg\vert_{\PP=0} \, ,
\end{split}
\eeq
where in the second line the gap perturbation is ${\redv(h(t))\to \redv(h(t)-\PP(t))}$ as mentioned earlier, corresponding to a perturbation ${P(t)=(\ell/d)\PP(t)}$ in the first line.

Note that we may express the isotropic response kernel through the same definition as in Ref.~\cite{AMZ19}, equivalent in high dimension, 
\beq
 M^{\text{iso}}_R(t,s)=  \frac\r d \sum_{\m=1}^d \int \de \rr_0 \, g_{\text{in}}(\rr_0) \, \frac{\d \moy{  \nabla_\m v\argp{\valabs{\rr_0'(t) + \vv(t)}} }_{\vv,\bm{P}}}{\d P_\m(s)}\Bigg\vert_{\bm P=\bm0}
 \, ,
\eeq
since we have (we start in the left-hand side from ${\bm P(s)=P(s)\hat\rr_0'(s)}$)
\beq
\begin{split}
 \frac{\d \moy{  \hat\rr_0'(t)\cdot\nabla v\argp{\valabs{\rr_0'(t) + \vv(t)}} }_{\vv,\bm{P}}}{\d P(s)}\Bigg\vert_{P=0}
 =&\sum_{\m=1}^d  \hat\rr_{0,\m}'(t)\frac{\d \moy{ \nabla_\m v\argp{\valabs{\rr_0'(t) + \vv(t)}} }_{\vv,\bm{P}}}{\d P(s)}\Bigg\vert_{P=0}\\
 =&\sum_{\m,\n}  \hat\rr_{0,\m}'(t)\frac{\partial P_\n(s)}{\partial P(s)}\frac{\d \moy{ \nabla_\m v\argp{\valabs{\rr_0'(t) + \vv(t)}} }_{\vv,\bm{P}}}{\d P_\n(s)}\Bigg\vert_{\bm P=\bm0}\\
  =&\sum_{\m,\n}  \hat\rr_{0,\m}'(t)\hat\rr_{0,\n}'(s)\frac{\d \moy{ \nabla_\m v\argp{\valabs{\rr_0'(t) + \vv(t)}} }_{\vv,\bm{P}}}{\d P_\n(s)}\Bigg\vert_{\bm P=\bm0}\\
  \sim&\sum_{\m=1}^d   \hat\rr_{0,\m}^2\frac{\d \moy{ \nabla_\m v\argp{\valabs{\rr_0'(t) + \vv(t)}} }_{\vv,\bm{P}}}{\d P_\m(s)}\Bigg\vert_{\bm P=\bm0}\\
  \sim&\sum_{\m=1}^d  \frac{\d \moy{ \nabla_\m v\argp{\valabs{\rr_0'(t) + \vv(t)}} }_{\vv,\bm{P}}}{\d P_\m(s)}\Bigg\vert_{\bm P=\bm0} \, ,
\end{split}
\eeq
where we used in the last two lines the symmetries, the fact that the shear-plane components are negligible in the sum in ${d\to\io}$, and that this {$\m$-$\m$} response is independent of $\m$ for $\m\geqslant3$.
In conclusion, the anisotropic components of ${M^{\m\n}_R(t,s)}$ are negligible for high $d$.

%-------
\subsection{Kernel ${\hat{k}(t)}$: average divergence of forces}
\label{sec-appendix-explicit-coeff-k-kappa}

The isotropic spring constant kernel, respectively bare and rescaled, is
\beq
\begin{split}
 k^{\text{iso}}(t)
 &	= \frac{1}{d} \sum_{\n=1}^d k^{\n\n}(t)
 	\stackrel{(d\to\io)}{=} \frac\r d \int \de \rr_0 \, g_{\text{in}}(\rr_0) \, \argc{ 	\moy{v''\argp{\valabs{\rr_0'(t)+\vv(t)}}}_{\vv}	
	+ d\moy{\frac{v'\argp{\valabs{\rr_0'(t)+\vv(t)}}}{\valabs{\rr_0'(t)+\vv(t)}}}_{\vv} 
	} \stackrel{(\m\neq1,2)}{=}k^{\m\m}(t)
 \, ,
 \\
  \kappa^{\text{iso}}(t)
 & \equiv \frac{\ell^2}{2 d^2} k^{\text{iso}}(t)
 	\stackrel{(d\to\io)}{=} \frac{\wh\f}{2} \int_{-\infty}^{\infty} \de h_0\, e^{h_0}  \int \mathcal{D} g_1 \mathcal{D} g_2
 \,g_{\text{in}}(h_0,g_1,g_2) \moy{\bar{v}'' (h(t))+ \bar{v}' (h(t))}_{h \vert h_0,g_1,g_2} \, .
\end{split}
\eeq
As for the other coefficients of the rescaled kernel, in the infinite-dimensional limit:
\beq
\begin{split}
 & \kappa^{\m\n}(t)
 \equiv \frac{\ell^2}{2d^2} k^{\m\n}(t)
 \\
  & \Rightarrow \quad \left\lbrace \begin{array}{rl}
 \kappa^{11}(t)
 	&=   \frac{\wh\f}{2} \int_{-\infty}^{\infty} \de h_0\, e^{h_0}  \int  \mathcal{D} g_1 \mathcal{D} g_2
	  \,g_{\text{in}}(h_0,g_1,g_2)\moy{\bar{v}'' (h(t)) \argp{g_1+\gamma(t)g_2}^2+ \bar{v}' (h(t))}_{h \vert h_0,g_1,g_2}
	   \, ,
 \\
 \kappa^{12}(t)&=
 	\kappa^{21}(t)=   \frac{\wh\f}{2} \int_{-\infty}^{\infty} \de h_0\, e^{h_0} \int  \mathcal{D} g_1 \mathcal{D} g_2
	  \, g_{\text{in}}(h_0,g_1,g_2)\moy{\bar{v}'' (h(t)) }_{h \vert h_0,g_1,g_2}\argp{g_1 g_2 + \gamma(t) g_2^2}
	   \, ,
 \\
 \kappa^{22}(t)&=
 	\frac{\wh\f}{2} \int_{-\infty}^{\infty} \de h_0\, e^{h_0} \int  \mathcal{D} g_1 \mathcal{D} g_2
	  \,g_{\text{in}}(h_0,g_1,g_2) \moy{\bar{v}'' (h(t)) }_{h \vert h_0,g_1,g_2}g_2^2
 \, , \\
  \kappa^{\mu\mu}(t)
   &= \kappa^{\text{iso}}(t) \quad \text{for }  \mu\geqslant3 \, .
 \end{array} \right.
\end{split}
\eeq
As for the force-force correlation, the components in the shear plane are of the same order of those for $\m\geqslant 3$, and are therefore negligible being much less numerous.

%_____________________________________________________________
\section{Projection of the vectorial effective stochastic process on the `longitudinal' motion}
\label{sec-projection-eff-stoch-process-detailed}

In this section we examine the projection of the fluctuating ${\vv(t)}$ on the drifted initial condition 
${\hat\rr_0'(t)\approx (\hat 1 +\hat \g(t))\hat \rr_0}$, providing the details that were skipped in Sec.~\ref{sec:scalar-formulation-eff-Langevin-process-longitud}.
We start from the effective vectorial stochastic process given in Eq.~\eqref{eq-coshearing-effective-stoch-processes}:
\beq
\label{eq:C1w}
\begin{split}
 & \frac{\z}{2} \dot \vv(t)
  = - \frac12 (\hat{k}(t)-\zeta \hat{\dot{\gamma}}(t)) \vv(t)
 	+ \frac12 \int_0^t \de s \, \hat{M}_R(t,s) \, \vv(s)
 	- \nabla v \argp{{\rr}_{0}'(t) + \vv(t)}
 	+ \bm{\Xi}(t)
 \, ,
 \\
 & \moy{\X_{\m}(t)}_{\bm{\Xi}}=0 \, , \quad
	\moy{\X_{\m}(t) \X_{\n}(s)}_{\bm{\Xi}}
	= \d_{\m\n} \left[ T \z \d(t-s) +\frac12 \G_C(t,s)\right] +\frac12 M^{\m\n}_C(t,s)-\bar\Xi_\m(t)\bar\Xi_\n(s)
 \, .
\end{split}
\eeq
As mentioned in Appendix~\ref{sec-appendix-explicit-coeff-overall-drift-Xi}, the last term in the variance can be neglected for ${d\to\io}$.

We start by examining each individual contributions.
Note that we underbrace the contributions stemming from the anisotropy under shear, which are thus simply absent in the derivation of Ref.~\cite{AMZ19}.
Moreover, anticipating that the statistical isotropy outside the shear plane would dominate in the infinite-dimensional limit, we separate the terms associated to the diagonal coefficients of the kernels.
We have first from the definition of the projected motion in Eq.~\eqref{def-longitud-projection-y-cs}:
\beq
 \textit{(i)} \quad
 	\frac{\zeta}{2}\frac{d}{\ell} \hat{\rr}_0'(t) \cdot \dot{\vv}(t)
 	= \frac\z2 \dot y(t)-\underbrace{\frac\z2 \g(t)\frac{g_2}{\sqrt d}\frac{d}{\ell}w_1(t)}_{\OO(\z/\sqrt d)} \, ,
\eeq
secondly
\beq
\begin{split}
 \textit{(ii)} \quad
 	- \frac12 \hat{\rr}_0'(t) \cdot \argp{\hat{k}(t) \vv(t)}\frac{d}{\ell}
 	= &	- \frac12 k^{\text{iso}}(t) \hat{\rr}_0'(t) \cdot \vv(t)\frac{d}{\ell}
	\\ 	
 	&	- \frac12 \argp{k^{11}(t) - k^{\text{iso}}(t)} \hat{r}_{0,1} w_{1}(t)\frac{d}{\ell}
		- \frac12 \argp{k^{22}(t) - k^{\text{iso}}(t)} \hat{r}_{0,2} w_{2}(t)\frac{d}{\ell}
 	\\
 	&	- \frac12 k^{12}(t) \hat{r}_{0,1} w_{2}(t)\frac{d}{\ell}
 		- \frac12 k^{21}(t) \hat{r}_{0,2} w_{1}(t)\frac{d}{\ell}
	\\
	\stackrel{(d\to\io)}{=} &	- \frac12 k^{\text{iso}}(t) y(t)+\OO\argp{\frac{k^{\text{iso}}}{\sqrt d}}
 \, ,
\end{split}
\eeq
thirdly
\beq
\begin{split}
 \textit{(iii)} \quad
 \frac\z2 \hat\rr_0'(t)\cdot\hat {\dot{\g}}(t)\vv(t)\frac{d}{\ell}=\frac{\zeta}{2} \dot{\gamma}(t) \frac{d}{\ell}w_{2}(t) \argp{\hat{r}_{0,1} + \gamma(t) \hat{r}_{0,2}} \stackrel{(d\to\io)}{=}\OO\argp{\frac{\z}{\sqrt d}}
 \, .
\end{split}
\eeq
Fourthly,  we showed in Appendix~\ref{sec-appendix-explicit-coeff-MR} that, with a slight abuse of 
notation\footnote{One should consider the projection of the original equation~\eqref{eqC3:GENLang-for-ucs-bis} 
instead of the projection of~\eqref{eq:C1w}.}, the memory kernel is effectively at leading order the projector 
${\hat M_R(t,s)\sim M_R^{\text{iso}}(t,s)\hat\rr_0'(t)\hat\rr_0'(s)^T}$. Consequently we get the leading order of the projected non-Markovian friction term
\beq
\begin{split}
 \textit{(iv)} \quad
 \frac12\frac{d}{\ell} \int_0^t \de s \, \hat{\rr}_0'(t) \argc{\hat{M}_R(t,s) \, \vv(s)} \stackrel{(d\to\io)}{=}&
  \frac12 \int_0^t \de s \, M_R^{\text{iso}}(t,s) \frac{d}{\ell}\hat\rr_0'(s)\cdot\vv(s)
 =  \frac12 \int_0^t \de s \, M_R^{\text{iso}}(t,s) \,y(s) \, .
\end{split}
\eeq
Fifthly the projected noise is Gaussian by linearity and has average 
${\moy{\hat{\rr}_0'(t) \cdot {\bf \Xi}(t)}=\hat{\rr}_0'(t) \cdot \moy{{\bf \Xi}(t)}=0}$ and variance
\beq
\begin{split}
 \textit{(v)}\quad \moy{\argc{\hat{\rr}_0'(t) \cdot {\bf \Xi}(t)}\argc{\hat{\rr}_0'(s) \cdot {\bf \Xi}(s)}}=
 \sum_{\m,\n}\hat r_{0,\m}'(t)\moy{\Xi_\m(t)\Xi_\n(s)}\hat r_{0,\n}'(s) \stackrel{(d\to\io)}{=}T\z\d(t-s)+\frac{\G_C(t,s)}{2}+\frac{M_C^{\text{iso}}(t,s)}{2}
 \, ,
\end{split}
\eeq
because the components in the shear plane are negligible in the latter sum. At last from Eq.~\eqref{eq-def-rescaled-potential-1}
\beq
\begin{split}
 \textit{(vi)} \quad
 - \hat{\rr}_0'(t) \cdot \nabla{v\argp{\valabs{ \rr_0'(t) + \vv(t)}}}
 \approx - v' \argp{{\rr}_{0}'(t) + \vv(t)} 
 \approx - \frac{d}{\ell} \bar{v}'(h(t))
 \, .
\end{split}
\eeq
Combining all these points, we obtain:
\beq
\begin{split}
 & \underbrace{\frac{\z \ell^2}{2d^2}}_{\equiv \widehat{\zeta}} \dot y(t)
  = - \underbrace{\frac{\ell^2}{2d^2} k^{\text{iso}}(t)}_{\equiv \kappa^{\text{iso}}(t) (t)}  y(t)
 	+  \int_0^t \de s \, \underbrace{\frac{\ell^2}{2d^2} M^{\text{iso}}_R(t,s)}_{\equiv \mathcal{M}^{\text{iso}}_R(t)(t,s)} \, y(s)
 	- \bar{v}(h(t))
 	+ \Xi(t)
 \, ,
 \\
 & \moy{\Xi(t)}_{\Xi}=0 \, , \quad
	\moy{\Xi(t) \Xi(s)}_{\Xi}
	= 2 T \hat{\zeta} \d(t-s) + \mathcal{G}_C(t,s) + \mathcal{M}^{\text{iso}}_C(t,s)
 \, .
\end{split}
\eeq
We finally redefine the kernels and the friction coefficient in order to reabsorbe the factors $d$ and $\ell$ (see Eq.~\eqref{eqC3:fricscal2-shear}), exactly as done in Ref.~\cite{AMZ19,KMZ16,MKZ16}, and this yields the effective stochastic process that characterises the fluctuations along the longitudinal direction ${\rr_0'(t)}$ (drifting with the strain) Eq.~\eqref{eq-def-longitudinal-motion-stoch-process}.

%_____________________________________________________________
%\bibliographystyle{mioaps}
%\bibliography{HS}

\end{document}